# Symmetry and magnitude of spin-orbit torques in ferromagnetic heterostructures

Kevin Garello[1], Ioan Mihai Miron[2], Can Onur Avci[1], Frank Freimuth[3], Yuriy Mokrousov[3], Stefan Blügel[3], Stéphane Auffret[2], Olivier Boulle[2], Gilles Gaudin[2], and Pietro Gambardella[1,4,5]

[1] *Catalan Institute of Nanotechnology (ICN), E-08193 Barcelona, Spain*

[2] *SPINTEC, UMR-8191, CEA/CNRS/UJF/GINP, INAC, F-38054 Grenoble, France*

[3] *Peter Grünberg Institut and Institute for Advanced Simulation, Forschungszentrum Jülich and JARA, 52425 Jülich, Germany*

[4] *Institució Catalana de Recerca i Estudis Avançats (ICREA), E-08010 Barcelona, Spain*

[5] *Department of Materials, ETH Zurich, CH-8093 Zurich, Switzerland*

**Current-induced spin torques are of great interest to manipulate the magnetization of thin films and nanostructures. Recent demonstrations of magnetization switching induced by in-plane current injection in heavy metal/ferromagnetic heterostructures have drawn attention to a class of spin torques based on orbital-to-spin momentum transfer, which is alternative to the spin transfer torque between noncollinear magnetic layers and amenable to more diversified device functions. The symmetry, magnitude, and origin of spin-orbit torques (SOTs), however, remain a matter of intense debate. Here we report on the three-dimensional vector measurement of SOTs in AlO$_x$/Co/Pt and MgO/CoFeB/Ta trilayers using harmonic analysis of the anomalous and planar Hall effects as a function of the applied current and magnetization direction. We provide an all-purpose scheme to measure the amplitude and direction of SOTs for any arbitrary orientation of the magnetization, including corrections due to the interplay of Hall and thermoelectric effects. Based on general space and time inversion symmetry arguments, we demonstrate that asymmetric heterostructures allow for two different SOTs having odd and even behavior with respect to magnetization reversal. Our measurements show that such torques include strongly anisotropic field-like and spin transfer-like components, which depend markedly on the annealing conditions and type of heavy metal layer. These results call for SOT models that go beyond the spin Hall and Rashba effects investigated thus far.**

Spintronic devices rely on the generation of spin torques to control the magnetization of miniaturized memory and logic elements using electric currents.[1,2] Conventionally, such torques have been associated with the transfer of spin angular momentum between a "polarizer" and a "free" ferromagnetic layer (FM) separated by a nonmagnetic spacer, mediated by a spin polarized current flowing perpendicular to the two layers.[2,3] In recent years, however, experiments[4-13] and theory[14-27] have pointed out alternative mechanisms to produce spin torques based on the spin-orbit interaction, which mediates the transfer of orbital angular momentum from the lattice to the spin system and do away with the need of a polarizer FM. These mechanisms, which include the spin Hall,[28] Rashba,[29] and Dresselhaus[30] effects, exploit the coupling between electron spin and orbital motion to induce nonequilibrium spin accumulation, which ultimately gives rise to a torque on the magnetization via the usual spin transfer channels between *s*- and *d*-electrons.[31,32] Henceforth, we refer to the torques induced by such effects as *spin-orbit torques* (SOTs) to underline their common link to the spin-orbit interaction.

Of particular relevance for magnetization switching, experiments on $AlO_x$/Co/Pt heterostructures have shown that current injection in the plane of the layers induces a spin accumulation component transverse to the current, $\delta \mathbf{m}^\perp \sim \mathbf{z} \times \mathbf{j}$,[5,6] as well as a longitudinal one that rotates with the magnetization in the plane defined by the current and the *z* axis of the stack, $\delta \mathbf{m}^\parallel \sim (\mathbf{z} \times \mathbf{j}) \times \mathbf{m}$,[9,33] where $\mathbf{j}$ and $\mathbf{m}$ are unit vectors that denote the current density and magnetization direction, respectively. Because of the exchange interaction between *s*- and *d*-electrons, these components produce two effective magnetic fields, $\mathbf{B}^\perp \sim \delta \mathbf{m}^\perp$ and $\mathbf{B}^\parallel \sim \delta \mathbf{m}^\parallel$, or, equivalently, a field-like torque $\mathbf{T}^\perp \sim \mathbf{m} \times \delta \mathbf{m}^\perp$ and a spin transfer-like torque $\mathbf{T}^\parallel \sim \mathbf{m} \times \delta \mathbf{m}^\parallel$. If $\mathbf{j}$ is injected along $\mathbf{x}$, these torques correspond to $\mathbf{T}^\perp \sim \mathbf{m} \times \mathbf{y}$ and $\mathbf{T}^\parallel \sim \mathbf{m} \times (\mathbf{y} \times \mathbf{m})$, respectively. Several studies have shown that $\mathbf{T}^\parallel$ is strong enough to reverse the magnetization of high-coercivity FM with both perpendicular[9,33,34] and in-plane[35] anisotropy for current densities of the order of $10^7 - 10^8$ A/cm$^2$.

The fact that the magnetization of a single-layer FM can be reversibly switched by lateral injection of an initially unpolarized current[9] has naturally elicited great interest in SOTs for technological applications. For example, $\mathbf{T}^{\parallel}$ has been proposed[9,36,37] and demonstrated[33] to induce switching of magnetic tunnel junction devices using a three-terminal configuration, where the read and write current paths are separated to avoid damage of the tunnel barrier. On the theoretical side, however, two apparently contrasting pictures have emerged: one based on the bulk spin Hall effect (SHE) in the heavy metal layer as the sole source of spin accumulation[9,20,23-25,33,35] and the other on Rashba-type effective fields and spin-dependent scattering effects that take place at the interface between the heavy metal and the FM.[9,21-26] Both pictures lead to qualitatively equivalent expressions for $\mathbf{T}^{\perp}$ and $\mathbf{T}^{\parallel}$ (Refs. 23-25) but differ in the relative magnitude of the torques, since a pure SHE implies $\mathbf{T}^{\parallel} \gg \mathbf{T}^{\perp}$ whereas the opposite is true if only interfacial Rashba contributions are considered.[25] On the experimental side, the situation is also controversial: In the same model system, $AlO_x$(1.6 nm)/Co(0.6 nm)/Pt(3 nm), measurements of $\mathbf{T}^{\perp}$ range from zero (Ref. 33) to 1 T/$10^8$ A/cm$^2$ (Ref. 5,6), whereas estimates of $\mathbf{T}^{\parallel}$ go from 0.017 (Ref. 33) to 0.08 T/$10^8$ A/cm$^2$ (Ref. 9). In MgO/CoFeB/Ta, Liu et al. reported that the direction of $\mathbf{T}^{\parallel}$ is reversed with respect to $AlO_x$/Co/Pt, consistently with the opposite sign of the spin Hall angle ($\theta_{SH}$) of Ta and Pt.[35] However, recent data show that the magnitude and even the sign of both $\mathbf{T}^{\parallel}$ and $\mathbf{T}^{\perp}$ depend on the thickness of the Ta layer,[38] highlighting how much these torques are sensitive to different effects. This state of affairs, which goes together with the lack of a consistent method of measuring SOTs of arbitrary orientation, makes it hard to optimize their efficacy for specific applications and reach a consensus on the physical origin of the torques.

The purpose of this paper is threefold. First, starting from symmetry arguments, we derive general expressions of the spin accumulation and current-induced SOTs in magnetic heterostructures that are independent of specific physical models. Second, we present a self-consistent, sensitive method to perform three-dimensional vector measurements of SOTs using an ac susceptibility technique based on the combination of the 1$^{st}$ and 2$^{nd}$ harmonic contributions of the anomalous Hall (AHE) and planar Hall effects (PHE). Third, we demonstrate unambiguously the existence of two distinct SOTs that have odd and even symmetry with respect to the inversion of the magnetization and include, but are not limited to, $\mathbf{T}^\perp \sim \mathbf{m} \times \mathbf{y}$ and $\mathbf{T}^\parallel \sim \mathbf{m} \times (\mathbf{y} \times \mathbf{m})$ (Fig. 1 a-c). We find strongly anisotropic SOT components that have not been observed thus far, which depend on the *x* and *y* projections of the magnetization in the plane of the current. $\mathbf{T}^\perp$ and $\mathbf{T}^\parallel$ have comparable magnitude in AlO$_x$/Co/Pt in the low current (low heating) limit and decrease significantly due to interface diffusion upon annealing. Both $\mathbf{T}^\perp$ and $\mathbf{T}^\parallel$ reverse sign and are dominated by anisotropy effects in MgO/CoFeB/Ta. The picture that emerges from this study is that, although SHE contributions to SOTs are important, interfacial effects play a prominent role in determining the magnitude and anisotropy of the torques.

**Spin-orbit torque symmetry and effective fields**

SOTs require inversion asymmetry in order to induce measurable effects on the magnetization, which is usually realized by sandwiching a FM between two different layers, as shown in Fig. 1. This is true also for torques produced by the SHE, which would average to zero in a symmetric heterostructure. The minimal requirements imposed by structure inversion asymmetry on a magnetic layer are continuous rotational symmetry around the *z* axis and mirror symmetry with respect to planes parallel to *z*. For this geometry we derived the general expressions for $\delta \mathbf{m}^{\perp,\parallel}$ and the resulting SOTs by expanding $\delta \mathbf{m}^{\perp,\parallel}$ in terms of the angular cosines of the magnetization defined with respect to the symmetry axis and current flow directions (see Supplementary Information). We find that the spin accumulation contains magnetization dependent terms that add to the $\delta \mathbf{m}^\perp \sim \mathbf{y}$ and $\delta \mathbf{m}^\parallel \sim \mathbf{y} \times \mathbf{m}$ components considered thus far, which change the symmetry and amplitude of $\mathbf{T}^\perp$ and $\mathbf{T}^\parallel$. The quantitative significance of these terms, however, must be established by experiment. Our measurements determine a minimal set of terms required to model the action of the field-like and spin transfer – like torques, namely

$$\mathbf{T}^\perp = (\mathbf{y} \times \mathbf{m})\,[T_0^\perp + T_2^\perp(\mathbf{z} \times \mathbf{m})^2 + T_4^\perp(\mathbf{z} \times \mathbf{m})^4] +$$
$$\mathbf{m} \times (\mathbf{z} \times \mathbf{m})(\mathbf{m} \cdot \mathbf{x})\,[T_2^\perp + T_4^\perp(\mathbf{z} \times \mathbf{m})^2] \qquad (1)$$

$$\mathbf{T}^\| = \mathbf{m} \times (\mathbf{y} \times \mathbf{m})\,T_0^\| + (\mathbf{z} \times \mathbf{m})(\mathbf{m} \cdot \mathbf{x})\,[T_2^\| + T_4^\|(\mathbf{z} \times \mathbf{m})^2]. \qquad (2)$$

Note that these torques are, respectively, odd and even with respect to the inversion of $\mathbf{m}$. For the special case $T_n^\perp = T_n^\| = 0$ for all $n \neq 0$, Eqs. 1 and 2 simplify to $\mathbf{T}^\perp = T_0^\perp(\mathbf{y} \times \mathbf{m})$ and $\mathbf{T}^\| = T_0^\|\,\mathbf{m} \times (\mathbf{y} \times \mathbf{m})$, which have been obtained theoretically for several models discussed in the literature.[21-26]

For the purpose of comparison with the experiment, we consider here the effective magnetic fields $\mathbf{B}^\perp$ and $\mathbf{B}^\|$ corresponding to $\mathbf{T}^\perp$ and $\mathbf{T}^\|$ obtained above. As the relevant field components are perpendicular to the magnetization, we adopt a spherical coordinate system (Fig. 1d) with basis vectors $\mathbf{e}_\theta = (\cos\theta \cos\varphi, \cos\theta \sin\varphi, -\sin\theta)$ and $\mathbf{e}_\varphi = (-\sin\varphi, \cos\varphi, 0)$, where $\mathbf{m} = (\sin\theta \cos\varphi, \sin\theta \sin\varphi, \cos\theta)$ and

$$\mathbf{B}^\perp = -\cos\theta \sin\varphi\,(T_0^\perp + T_2^\perp \sin^2\theta + T_4^\perp \sin^4\theta)\mathbf{e}_\theta - \cos\varphi\,T_0^\perp \mathbf{e}_\varphi, \qquad (3)$$

$$\mathbf{B}^\| = \cos\varphi\,(T_0^\| + T_2^\| \sin^2\theta + T_4^\| \sin^4\theta)\mathbf{e}_\theta - \cos\theta \sin\varphi\,T_0^\| \mathbf{e}_\varphi. \qquad (4)$$

Using Eqs. (3) and (4), the action of the current-induced fields on the magnetization can be directly compared to that of a reference external field ($\mathbf{B}_{\text{ext}}$) of known magnitude and direction by means of low frequency susceptibility measurements.[6,39]

**Hall measurements of current-induced effective fields**

To quantify and analyze the SOT fields described above, we studied $AlO_x$(2 nm)/Co(0.6 nm)/Pt(3 nm) as model system, deposited on oxidized Si by dc-magnetron sputtering. The $AlO_x$/Co stack was patterned into 1 x 1 and 1 x 0.5 µm² rectangular dots and the Pt layer etched into a cross, as shown in Fig. 1, which allows for current injection and Hall voltage ($V_H$) detection. We used an ac current of frequency $f$ to modulate the SOT amplitude and induce small oscillations of **m** about its equilibrium direction, defined by $\mathbf{B}_{ext}$ and the magnetic anisotropy of the trilayer. Such oscillations generate a second harmonic contribution to $V_H$, which provides a sensitive way to measure current-induced fields (Supplementary Information). In general, $V_H$ depends on $m_z$ through the AHE and on the product $m_x m_y$ through the PHE:

$$V_H = R_{AHE} I \cos\theta + R_{PHE} I \sin^2\theta \sin 2\varphi, \qquad (5)$$

where $R_{AHE}$ and $R_{PHE}$ are the AHE and PHE resistances, respectively, and $I$ is the injected current. In terms of the total Hall resistance $R_H = V_H/I$, the first harmonic term $R_H^f = R_{AHE}^f + R_{PHE}^f$ relates to the equilibrium direction of the magnetization and is independent of modulated fields. The second harmonic term $R_H^{2f}$ measures the susceptibility of the magnetization to the current-induced fields and is given by:

$$R_H^{2f} = (R_{AHE} - 2R_{PHE}\cos\theta \sin 2\varphi)\frac{d\cos\theta}{dB_{ext}}\frac{B_\theta}{\sin(\theta_B - \theta)} + 2R_{PHE}\sin^2\theta \cos 2\varphi \frac{B_\varphi}{B_{ext}\sin\theta_B}, \qquad (6)$$

where $B_\theta$ and $B_\varphi$ represent the polar and azimuthal components of the total effective field $\mathbf{B}^\perp + \mathbf{B}^\parallel$ induced by the current and $\theta_B$ is the polar angle of $\mathbf{B}_{ext}$ (Fig. 1d). Equation 6 allows us to measure $B_\theta$ and $B_\varphi$ as a function of $\theta$ and $\varphi$ (see Supplementary Information for more details). If $R_{PHE} = 0$, it is straightforward to evaluate $B_\theta$ from Eq. 6 by noting that $R_{AHE}\frac{d\cos\theta}{dB_{ext}} = \frac{dR_H^f}{dB_{ext}}$. Otherwise, $B_\theta$ and $B_\varphi$ must be evaluated by measuring $V_H$ at $\varphi = 0°$ and 90° and fitting $R_H^{2f}$ using a recursive procedure that accounts for both the AHE and the PHE ($R_{AHE}$ = 0.72 Ω and $R_{PHE}$ = 0.09 Ω for the sample presented in Figs. 1-4). This evaluation procedure has been validated by numerical macrospin simulations as well as by application of external ac fields in phase and antiphase with the current, the amplitude of which was correctly retrieved using Eq. 6 (see Supplementary Information). For full characterization, the sample was mounted on a rotating stage and $\mathbf{B}_{ext}$ applied in different directions, defined by polar ($\theta_B$) and azimuthal ($\varphi_B = \varphi$) coordinates. The current was modulated at a frequency $f$ = 10Hz. At each field point, $V_H$ has been measured during 10 s and fast Fourier transformed to extract $R_H^f$ and $R_H^{2f}$. The ac current amplitude was varied up to $I$ = 1136 µA, corresponding to a current density $j$ = 3.15 x $10^7$A/cm².

Figures 1e and f show $R_H^f$ measured as a function of $\mathbf{B}_{ext}$ applied out-of-plane ($\theta_B$ = 0°) and nearly in-plane ($\theta_B$ = 82°), respectively. The curves, proportional to $m_z$, are characteristic of AlO$_x$/Co/Pt layers with strong PMA. The slow and reversible decrease of $R_H^f$ with increasing in-plane field observed in Fig. 1f is due to the coherent rotation of the Co magnetization towards the hard plane direction. Figure 2 shows the second harmonic measurements of $R_H^{2f}$ for $\mathbf{B}_{ext}$ applied at $\theta_B$ = 82°, perpendicular ($\varphi$ = 90°, Fig.2a) and parallel ($\varphi$ = 0°, Fig.2b) to the current. The data are shown after subtraction of sample-dependent contributions to the Hall voltage that are not included in Eq. 5, namely a constant offset due to the voltage probe asymmetry as well as the anomalous Nernst-Ettinghausen effect (ANE). The ANE can be separately measured, giving a small correction to $R_H^{2f}$ of the order of 0.1 mΩ (Supplementary Information). We note that the choice of $\theta_B$ is not critical as long as $\mathbf{B}_{ext}$ is slightly tilted off-plane, to prevent the formation of magnetic domains. According to Eq. 6, $R_H^{2f}$ is mostly sensitive to the effective field components parallel to $\mathbf{e}_\theta$, as these affect $m_z$ and hence the AHE. Conversely, the components parallel to $\mathbf{e}_\varphi$ are measured through the PHE, which is significantly weaker. Thus, $R_H^{2f}$ measured at $\varphi$ = 90° reflects mostly $\mathbf{B}^\perp$ contributions, whereas $R_H^{2f}$ measured at $\varphi$ = 0° reflects mostly $\mathbf{B}^\parallel$ terms. Note that this agrees with the even/odd character of $R_H^{2f}$ measured at $\varphi = 90°/0°$ with respect to field inversion (Fig. 2a and b), since $\mathbf{B}^\perp$ and $\mathbf{B}^\parallel$ are by definition even and odd with respect to $\mathbf{m}$, opposite to the torques from which they are derived.

**Field-like and spin transfer-like torque components**

The effective fields $\mathbf{B}^\perp$ and $\mathbf{B}^\parallel$ derived from $R_H^{2f}$ are shown in Fig. 2c and d for $\mathbf{m}$ // $\mathbf{y}$ ($\varphi = 90°$) and $\mathbf{m}$ // $\mathbf{x}$ ($\varphi = 0°$), respectively. These measurements represent a central result of this work, providing a direct quantitative estimate of the SOT fields for different orientations of the magnetization. We find several interesting features that reveal a more complex scenario than previously anticipated. In particular, $\mathbf{B}^\perp$ depends not only on the current amplitude but also on the applied magnetic field, with a minimum of about 10 mT at $B_{ext} = 0$ T and increasing up to 20 mT for $B_{ext} \geq 1$ T. Since the external field only changes the orientation of the magnetization, this behaviour indicates that $\mathbf{B}^\perp$ depends strongly on the direction of $\mathbf{m}$. By converting the field dependence into a $\theta$ dependence using the AHE, we find that $B^\perp$ measured at $\varphi = 90°$ closely follows the function $-\cos\theta \, (T_0^\perp + T_2^\perp \sin^2\theta)$ with $T_0^\perp = -11.6 \pm 0.7$ mT and $T_2^\perp = -11.2 \pm 0.6$ mT, as shown in Fig. 2e. This expression is in agreement with Eq. 3, but differs remarkably from that expected from either the Rashba field[5,14,24-26] or the field-like component of the SHE torque[23,25] reported in the literature, which would require $T_2^\perp = 0$. We note that $T_0^\perp$ includes the contribution of the Oersted field produced by the current flowing in the Pt layer, which we estimate as $\frac{\mu_0 I}{2L} = -0.7$ mT (antiparallel to $\mathbf{y}$), where $L$ is the width of the current line and $\mu_0$ the vacuum permeability.

The dependence of $\mathbf{B}^\parallel$ on the magnetization is remarkably different from $\mathbf{B}^\perp$. Figure 2f shows that $B^\parallel$ measured at $\varphi = 0°$ is weakly dependent on $\theta$, and is well-approximated by $T_0^\parallel + T_2^\parallel \sin^2\theta + T_4^\parallel \sin^4\theta$, in agreement with Eq. 4, with $T_0^\parallel = 19.0 \pm$ 0.5 mT, $T_2^{\parallel} = 2 \pm 1$ mT, and $T_4^{\parallel} = -1 \pm 1$ mT. As the higher order coefficients are small and tend to compensate, $B^{\parallel}$ can be reasonably approximated by a constant value $T_0^{\parallel}$, consistently with previous findings.[9,33,34] This behavior is typical of all our AlO$_x$/Co/Pt samples, apart from small changes of the coefficients that we attribute to pattern or material inhomogeneities.

In order to complete the description of $\mathbf{B}^{\perp}$ and $\mathbf{B}^{\parallel}$, we performed a series of measurements for different orientations of $\mathbf{m}(\theta, 0° \leq \varphi \leq 90°)$. In such a case, $R_H^{2f}$, shown in Fig. 3a, is given by the linear superposition of two terms $R_H^{2f}(\mathbf{B}^{\perp}) + R_H^{2f}(\mathbf{B}^{\parallel})$, which can be easily separated owing to their different symmetry with respect to the inversion of $\mathbf{m}$ (see Supplementary Information). Figures 3b and c show $R_H^{2f}(\mathbf{B}^{\perp})$ and $R_H^{2f}(\mathbf{B}^{\parallel})$, respectively, as a function of $\varphi$. The lineshape of $R_H^{2f}(\mathbf{B}^{\perp})$ is similar to $R_H^{2f}$ measured at $\varphi = 90°$, shown in Fig. 2a, whereas $R_H^{2f}(\mathbf{B}^{\parallel})$ is similar to $R_H^{2f}$ measured at $\varphi = 0°$, shown in Fig. 2b. However, the amplitude of $R_H^{2f}(\mathbf{B}^{\perp})$ increases whereas $R_H^{2f}(\mathbf{B}^{\parallel})$ decreases as $\varphi$ goes from 0° to 90°. From these curves we obtain that the polar component of $\mathbf{B}^{\perp}$ scales proportionally to $\sin\varphi$, whereas the polar component of $\mathbf{B}^{\parallel}$ scales as $\cos\varphi$, in agreement with Eqs. (3) and (4), respectively. This implies that, within the error of our data, the SOT coefficients $T_0^{\perp}$, $T_2^{\perp}$, and $T_0^{\parallel}$ are independent of $\varphi$ (Fig. 3d-f), in agreement with the superposition principle for the current and the resulting linear-response torques.

**Torque to current ratios**

Figure 4 shows that the amplitude of $\mathbf{T}^{\perp}$ and $\mathbf{T}^{\parallel}$ scales linearly with the current up to $j = 1.5 \times 10^7$ A/cm$^2$. Above this value, we observe a nonlinear increase of the coefficients $T_0^{\perp}$, $T_2^{\perp}$, and $T_0^{\parallel}$, which we attribute to Joule heating effects. We note that, at the maximum current density employed in this study (3.15 x 10$^7$ A/cm$^2$), we observe a reduction of the AHE (-3.5%) and magnetic anisotropy (-13%) as well as an increase of the resistivity of the layers (+13%). All these effects can be attributed to heating and may alter the SOT/current ratio.

Our measurements also offer quantitative insight into the magnitude of the different kinds of SOTs. In the following, we express the torques per unit of magnetic moment, thereby assigning them the same units as the effective fields. We discuss first $\mathbf{T}^\perp$. From the initial slope of the data reported in Fig. 4, we find that the SOT/current ratios corresponding to $T_0^\perp$ and $T_2^\perp$ are -3.2 ± 0.2 and -2.3 ± 0.2 mT per $10^7$ A/cm$^2$, respectively. This corrects our previous estimate of $\mathbf{T}^\perp$ based on current-induced domain nucleation,[5] which largely overestimated $T_0^\perp$ due to heat-assisted magnetization reversal[6] and neglect of $T_2^\perp$. Moreover, we prove beyond doubt that $\mathbf{T}^\perp$ is not an apparent effect due to spin Hall torque dynamics in the high current regime. This hypothesis was suggested by Liu et al., who reported no evidence of $\mathbf{T}^\perp$ in AlO$_x$/Co/Pt within a sensitivity of 1.3 mT per $10^7$ A/cm$^2$ (Ref. 33). Our measurements, however, are quasi-static and extend well into the low current regime. We suggest that the negative result of Liu et al. might be partly due to the different preparation of the AlO$_x$/Co/Pt stack (oxidized in air and annealed up to 350ºC) and partly to the lower sensitivity afforded by the dc measurements reported in Ref. 33 (a comparison of ac and dc measurements is reported in the Supplementary Information).

By fitting $T_0^\parallel$ at low current density ($j < 1.5$ x $10^7$ A/cm$^2$, Fig. 4c), we obtain $T_0^\parallel$ = 5.0 ± 0.2 mT per $10^7$ A/cm$^2$. Note, however, that the data reported in Fig. 4 (dots and squares) represent a lower bound of the torque/current ratio due to current dispersion in the Hall voltage probes, which reaches up to 23% of the total current in a square Hall cross.[40] Measurements of Hall crosses with narrower voltage probes (0.5 μm instead of 1 μm) give consistently higher torque/current ratios, namely $T_0^\perp$ = -4.0 ± 0.3, $T_2^\perp$ = -2.7 ± 0.1, and $T_0^\parallel$ = 6.9 ± 0.3 mT per $10^7$ A/cm$^2$ (Fig. 4, triangles, and Supplementary

Information). This value of $T_0^\parallel$ is about four times larger than that reported by Liu et al. for AlO$_x$/Co/Pt (1.7 mT per $10^7$ A/cm$^2$),[33] and also larger compared to MgO/CoFeB/Ta (4 mT per $10^7$ A/cm$^2$).[35] Supposing that **T**$^\parallel$ originates uniquely from the bulk spin Hall effect in Pt and that the Pt/Co interface is fully transparent, implies $T_0^\parallel = \frac{\hbar}{2eMt_{Co}} \vartheta_{SH} j \left[1 - \text{sech}(t_{Pt}/\lambda_{Pt})\right]$, where $e$ is the electron charge, $M = 10^6$ A/m, $t_{Co}$ = 0.6 nm, and $\vartheta_{SH}$ is the spin Hall angle of Pt. The last factor takes into account the partial compensation of the spin Hall current due to spin diffusion from the bottom Pt interface, which depends on the spin diffusion length of Pt ($\lambda_{Pt}$).[33] By assuming $\lambda_{Pt}$ = 1.4 nm,[41] our measurement of $T_0^\parallel$ implies $\vartheta_{SH} = 0.16$, which is comparable to Ta[35] and more than twice the largest spin Hall angle reported previously for Pt.[41] Taking $\lambda_{Pt}$ = 10 nm, as discussed in a recent report,[42] would push $\vartheta_{SH}$ to even higher values. We conclude that either $\vartheta_{SH}$ is much larger than expected or additional spin-orbit and interface-related effects contribute to **T**$^\parallel$.

**Torque dependence on interface and material parameters**

To investigate how the SOTs depend on the quality of the AlO$_x$/Co/Pt interfaces, we measured **B**$^\perp$ and **B**$^\parallel$ on trilayers annealed to 300 °C for 30 minutes in vacuum. We find that annealing induces a significant degradation of the SOT amplitude, corresponding to a reduction of $T_0^\perp$, $T_2^\perp$, and $T_0^\parallel$ by about 17 %, 60%, and 23%, respectively (Fig. 5a and b). The resistivity, which is 36 μΩcm in the as deposited samples, increases by about 7%, whereas the AHE goes from 0.80 to 1.14 Ω. This is consistent with previous measurements of annealed AlO$_x$/Co/Pt trilayers,[43] where the AHE increase was attributed to the diffusion of Pt atoms into the Co layer. Since annealing above 250 °C is known to induce mixing of Co and Pt and affect the oxidation of the AlO$_x$/Co interface,[43,44] we conclude that both **T**$^\perp$ and **T**$^\parallel$ are very sensitive to the interface quality of the trilayers.

Figure 5c and d show $\mathbf{B}^\perp$ and $\mathbf{B}^\parallel$ measured on MgO(2 nm)/CoFeB(0.9 nm)/Ta(3 nm) layers annealed to 300 °C in vacuum. By comparison with Fig. 5a and b, it is evident that $\mathbf{B}^\perp$ and $\mathbf{B}^\parallel$ reverse sign compared to AlO$_x$/Co/Pt, consistently with previous studies.[35,38] However, the strong θ dependence of both fields, not observed before, reveals that the SOT anisotropy is a general effect that is not unique to AlO$_x$/Co/Pt. The fits of $\mathbf{B}^\perp$ and $\mathbf{B}^\parallel$ according to Eqs. 3 and 4 give $T^\perp_{0,2,4}$= 4.5±0.1, 5.6±0.2, 5.9±0.3 and $T^\parallel_{0,2,4}$ = -2.4±0.1, 0.4±0.4, -2.0±0.4 mT per $10^7$ A/cm$^2$. Thus, both 2$^{nd}$ and 4$^{th}$ order terms of amplitude comparable to the 0$^{th}$ order are required to model the angular dependence of $\mathbf{B}^\perp$ and $\mathbf{B}^\parallel$ in MgO/CoFeB/Ta. We note also that our value of $T^\parallel_0$ is in between that reported by Refs. 35,38, whereas the field-like terms are considerably larger than the spin-transfer like ones, contrary to what is found for perpendicular current injection in metallic spin valve systems.[45]

In conclusion, general symmetry arguments show that $\mathbf{T}^\perp$ and $\mathbf{T}^\parallel$ can have a complex vector dependence on the direction of the magnetization. This work provides the first evidence for this effect as well as a method to measure $\mathbf{T}^\perp$, $\mathbf{T}^\parallel$, and their dependence on the magnetization in vector form. We find that there are significant deviations from the SOT models considered so far based on the Rashba and spin Hall effects. In the case of AlO$_x$/Co/Pt, the largest deviations are observed for $\mathbf{T}^\perp$ due to terms proportional to $T^\perp_2$ in Eq. 1. Thus, the effective field $\mathbf{B}^\perp$ generated by the current includes magnetization-dependent components perpendicular to the *y* axis, whereas the Rashba model can only explain components parallel to *y*. This suggests that the previous picture of $\mathbf{T}^\perp$ induced by a Rashba field of constant magnitude[9,21-26] or the field-like component of the spin Hall torque[23,25] has to be extended by a calculation of the torque based on a realistic description of the electronic structure. In the case of MgO/CoFeB/Ta, both $\mathbf{T}^\perp$ and $\mathbf{T}^\parallel$ present strong anisotropic components, which maximize the torques when the magnetization lies in the plane of the FM. Tuning of the vector properties of SOTs may play a crucial role in developing novel spintronic devices where different magnetic states are induced by specific SOT components.

**METHODS**

The samples were fabricated from Al(1.6 nm)/Co(0.6 nm)/Pt(3 nm) and MgO(2)/CoFeB(0.9)/Ta(3) layers deposited on a thermally oxidized silicon wafer by dc magnetron sputtering. The deposition rates were 0.05 nm/s (Co and Al), 0.15 nm/s (Ta) and 0.1 nm/s (Pt, Mg) at an Ar pressure of $2 \times 10^{-3}$ mbar. After deposition the Al/Co/Pt films were oxidized by exposure to a radiofrequency (rf) oxygen plasma at a pressure of $3 \times 10^{-3}$ mbar and an rf power of 10 W for 29 s. Mg/CoFeB/Ta was naturally oxidized in an oxygen pressure of 150 mbar for 10 s. The $AlO_x$/Co/Pt films were patterned by e-beam lithography and ion beam etching into 1000 and 500 nm square $AlO_x$/Co dots and Pt Hall crosses. The typical resistance of these devices is 3-4 k$\Omega$ and is mostly due to the thin Pt contact leads, whereas the resistivity of $AlO_x$/Co/Pt is 36 $\mu\Omega$cm. The MgO/CoFeB/Ta layers were patterned into 1000 nm wide Hall bars with 500 nm voltage branches. The resistivity of our MgO/CoFeB/Ta devices is 184 $\mu\Omega$cm. The Hall voltage measurements were performed at room temperature by using an ac current of amplitude 200 to 1136 $\mu$A modulated at $f = 10$ Hz.

# References

**Acknowledgements** This work was supported by the European Research Council (StG 203239 NOMAD), the European Commission under the Seventh Framework Programme (GA 318144, SPOT), Ministerio de Economía y Competitividad (ERA-Net EUI2008-03884, MAT2010-15659), Agència de Gestió d'Ajuts Universitaris i de Recerca (2009 SGR 695), and the Agence Nationale de la Recherche (ANR-10-BLANC-1011-3 "SPINHALL"). F.F. and Y.M. acknowledge funding under the HGF-YIG program VH-NG-513. The samples were patterned at the NANOFAB facility of the Institut Néel (CNRS).

**Author contributions** K.G., I.M.M., and P.G. planned the experiment; I.M.M., C.O.A., G.G., and S.A. fabricated the samples. K.G., I.M.M., and C.O.A. performed the measurements; K.G., I.M.M., C.O.A., and P.G. analyzed the data; F.F. derived the general expression for the torques, K.G. and P.G. wrote the manuscript. All authors discussed the results and commented on the manuscript.

**Author Information**
The authors declare no competing financial interests. Correspondence and requests for materials should be addressed to K.G. (kevin.garello@icn.cat) and P.G. (pietro.gambardella@mat.ethz.ch).

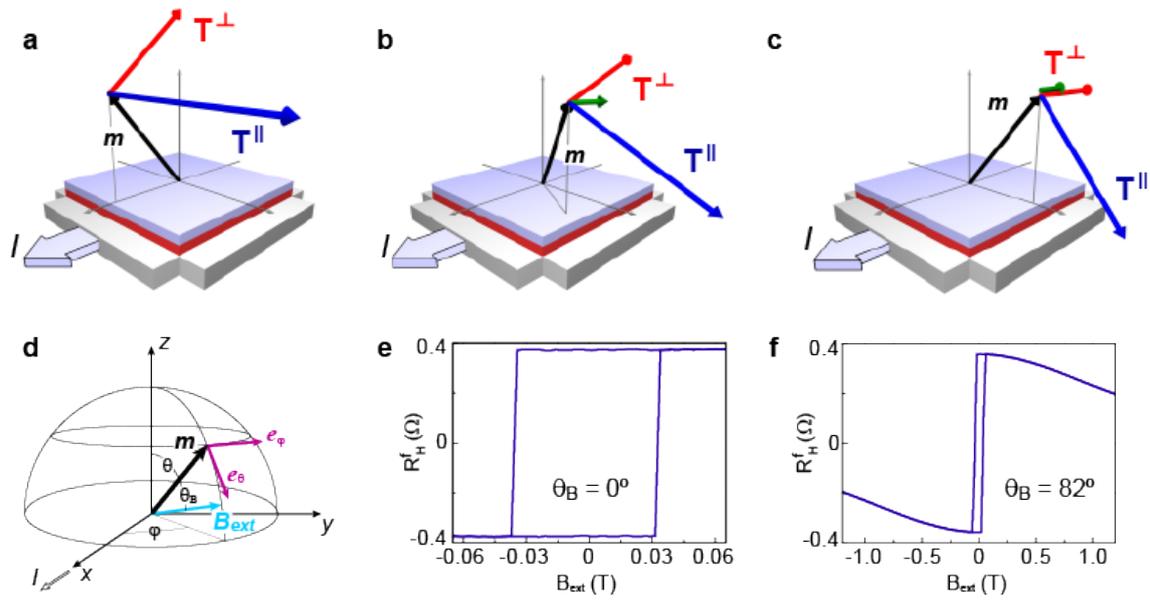

**Figure 1**. **Torque schematics and magnetization measurements**. $AlO_x$/Co/Pt Hall cross with current and voltage leads. The arrows indicate the direction and amplitude of the torques for **a,** ($\varphi = 0°$), **b,** ($\varphi = 60°$), and **c,** ($\varphi = 90°$). The field-like and anisotropic components of $\mathbf{T}^{\perp}$ are shown in red and green, respectively. $\mathbf{T}^{\parallel}$ is shown in blue. **d,** Coordinate system. **e,** $m_z$ measured by the first harmonic Hall resistance $R_H^f$ as a function of applied field $B_{ext}$ parallel to the easy axis ($\theta_B = 0°$) and **f,** close to in-plane ($\theta_B = 82°$, $\varphi = 0°$).

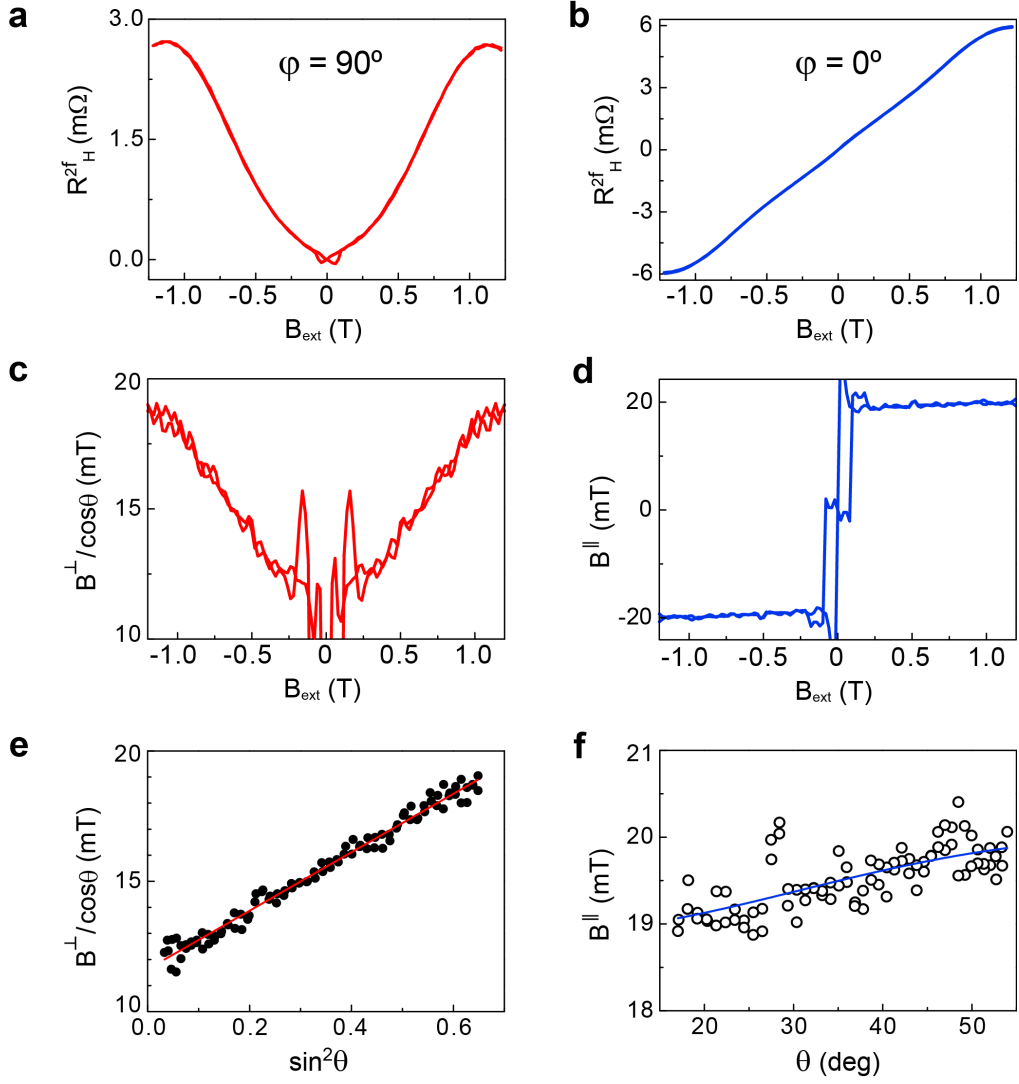

**Figure 2. Second harmonic Hall resistance and current-induced spin-orbit fields. a,** $R_H^{2f}$ measured as a function of $B_{ext}$ applied at $\theta_B = 82°$, $\varphi = 90°$ and **b,** $\theta_B = 82°$, $\varphi = 0°$. The amplitude of the ac current is 1.136 mA. **c,** Effective field $B^\perp/\cos\theta$ measured at $\varphi = 90°$ as a function of $B_{ext}$. **d,** Effective field $B^\parallel$ measured at $\varphi = 0°$ as a function of $B_{ext}$. **e,** $B^\perp/\cos\theta$ measured at $\varphi = 90°$ as a function of $\sin^2\theta$. The solid line is a fit to $T_0^\perp + T_2^\perp \sin^2\theta$ according to Eq. (3). **f,** $B^\parallel$ measured at $\varphi = 0°$ as a function of $\theta$. The solid line is a fit to $T_0^\parallel + T_2^\parallel \sin^2\theta + T_4^\parallel \sin^4\theta$ according to Eq. (4). Note that, in our units, $|\mathbf{T}^\perp| = |B^\perp/\cos\theta|$ for $\varphi = 90°$ and $|\mathbf{T}^\parallel| = |B^\parallel|$ for $\varphi = 0°$.

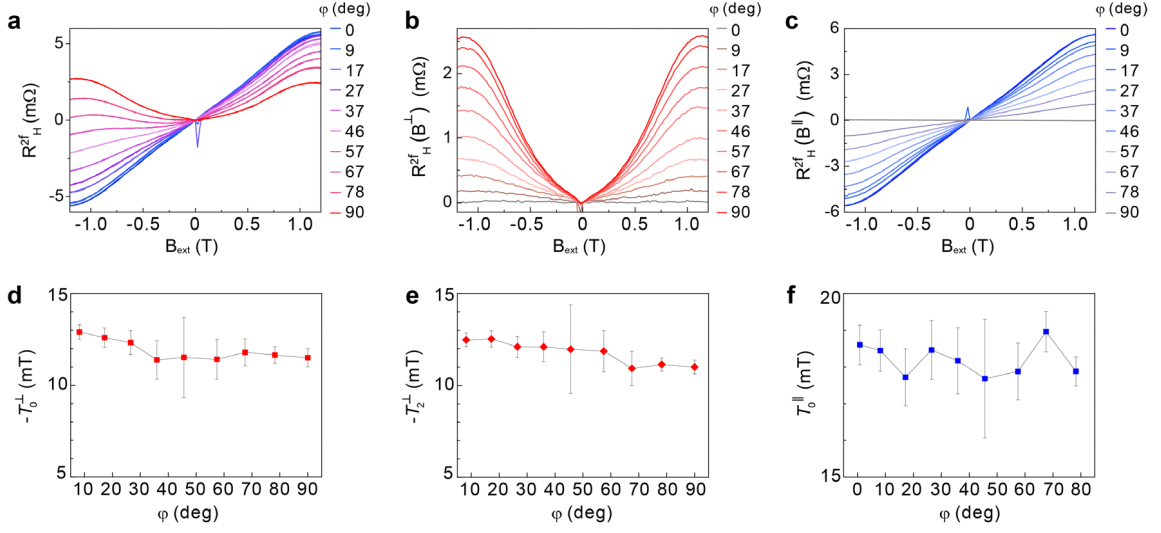

**Figure 3. Angular dependence of the Hall resistance and SOT coefficients. a,** $R_H^{2f}$ as a function of $B_{ext}$ applied at $\theta_B = 82°$ measured for different in-plane orientations of the magnetization. **b,** Symmetric $R_H^{2f}(\mathbf{B}^\perp)$ and **c,** antisymmetric $R_H^{2f}(\mathbf{B}^\parallel)$ components of $R_H^{2f}$. SOT coefficients **d,** $T_0^\perp$, **e,** $T_2^\perp$, **f,** $T_0^\parallel$ as a function of $\varphi$. The amplitude of the ac current is 1.136 mA.

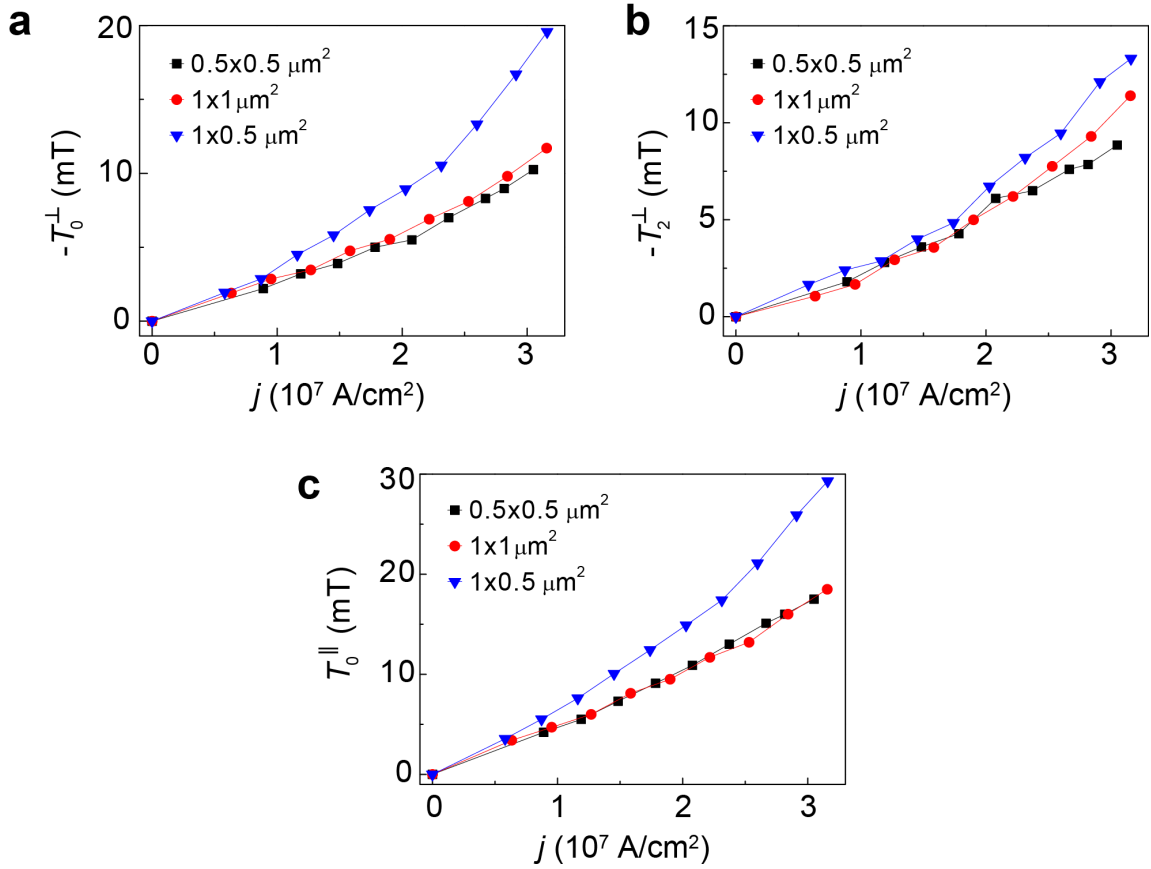

**Figure 4. Dependence of the SOT coefficients on the injected current density. a,** $T_0^\perp$, **b,** $T_2^\perp$, **c,** $T_0^\parallel$ as a function of *j* for different samples. Red dots and black squares refer to symmetric Hall crosses with 1x1 and 0.5x0.5 μm² dimensions, respectively. The blue triangles refer to an asymmetric Hall cross with a 1 μm wide current line and 0.5 μm wide voltage probes.

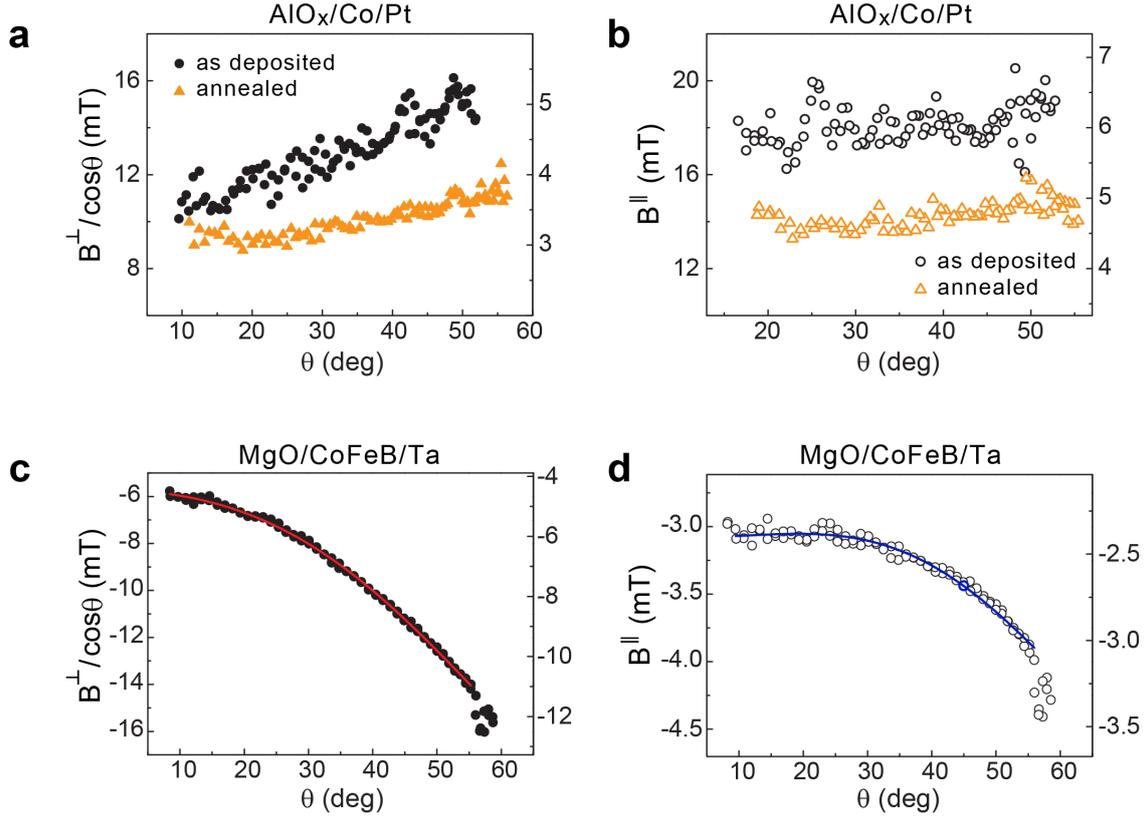

**Figure 5. Effect of thermal annealing and material composition on the current-induced spin-orbit fields. a,** Effective field $B^\perp/\cos\theta$ measured at $\varphi = 90°$ and **b,** effective field $B^\parallel$ measured at $\varphi = 0°$ in AlOx/Co/Pt as a function of $\theta$. The measurements refer to a 0.5x0.5 µm² as deposited sample (black dots) annealed to 300°C (green triangles). The amplitude of the ac current is 540 µA and 550 µA for the as deposited and annealed samples, respectively. **c,** $B^\perp/\cos\theta$ measured at $\varphi = 90°$ and **d,** $B^\parallel$ measured at $\varphi = 0°$ for an MgO/CoFeB/Ta Hall bar annealed to 300°C. The width of the Hall bar is 1 µm and the amplitude of the ac current is 500 µA. The scale on the right hand side of the plots is in mT per $10^7$ A/cm².

1   Chappert, C., Fert, A. & Van Dau, F. N. The emergence of spin electronics in data storage. *Nat. Mater.* **6**, 813-823, (2007).

2   Brataas, A., Kent, A. D. & Ohno, H. Current-induced torques in magnetic materials. *Nat. Mater.* **11**, 372-381, (2012).

3   Ralph, D. C. & Stiles, M. D. Spin transfer torques. *J. Magn. Magn. Mater.* **320**, 1190-1216, (2008).

4   Chernyshov, A., Overby, M., Liu, X. Y., Furdyna, J. K., Lyanda-Geller, Y. & Rokhinson, L. P. Evidence for reversible control of magnetization in a ferromagnetic material by means of spin-orbit magnetic field. *Nat. Phys.* **5**, 656-659, (2009).

5   Miron, I. M., Gaudin, G., Auffret, S., Rodmacq, B., Schuhl, A., Pizzini, S., Vogel, J. & Gambardella, P. Current-driven spin torque induced by the Rashba effect in a ferromagnetic metal layer. *Nat. Mater.* **9**, 230-234, (2010).

6   Pi, U. H., Won Kim, K., Bae, J. Y., Lee, S. C., Cho, Y. J., Kim, K. S. & Seo, S. Tilting of the spin orientation induced by Rashba effect in ferromagnetic metal layer. *Appl. Phys. Lett.* **97**, 162507, (2010).

7   Fang, D., Kurebayashi, H., Wunderlich, J., Vyborny, K., Zarbo, L. P., Campion, R. P., Casiraghi, A., Gallagher, B. L., Jungwirth, T. & Ferguson, A. J. Spin-orbit-driven ferromagnetic resonance. *Nat Nanotechnol* **6**, 413-417, (2011).

8   Suzuki, T., Fukami, S., Ishiwata, N., Yamanouchi, M., Ikeda, S., Kasai, N. & Ohno, H. Current-induced effective field in perpendicularly magnetized Ta/CoFeB/MgO wire. *Appl. Phys. Lett.* **98**, 142505, (2011).

9   Miron, I. M., Garello, K., Gaudin, G., Zermatten, P. J., Costache, M. V., Auffret, S., Bandiera, S., Rodmacq, B., Schuhl, A. & Gambardella, P. Perpendicular switching of a single ferromagnetic layer induced by in-plane current injection. *Nature* **476**, 189-193, (2011).

10  Kajiwara, Y., Harii, K., Takahashi, S., Ohe, J., Uchida, K., Mizuguchi, M., Umezawa, H., Kawai, H., Ando, K., Takanashi, K., Maekawa, S. & Saitoh, E. Transmission of electrical signals by spin-wave interconversion in a magnetic insulator. *Nature* **464**, 262-266, (2010).

11  Kurebayashi, H., Dzyapko, O., Demidov, V. E., Fang, D., Ferguson, A. J. & Demokritov, S. O. Controlled enhancement of spin-current emission by three-magnon splitting. *Nat. Mater.* **10**, 660-664, (2011).

12  Liu, L., Moriyama, T., Ralph, D. C. & Buhrman, R. A. Spin-Torque Ferromagnetic Resonance Induced by the Spin Hall Effect. *Phys. Rev. Lett.* **106**, 036601 (2011).

13	Demidov, V. E., Urazhdin, S., Ulrichs, H., Tiberkevich, V., Slavin, A., Baither, D., Schmitz, G. & Demokritov, S. O. Magnetic nano-oscillator driven by pure spin current. *Nat. Mater.* **11**, 1028-1031, (2012).

14	Manchon, A. & Zhang, S. Theory of nonequilibrium intrinsic spin torque in a single nanomagnet. *Phys. Rev. B* **78**, 212405 (2008).

15	Obata, K. & Tatara, G. Current-induced domain wall motion in Rashba spin-orbit system. *Phys. Rev. B* **77**, 214429 (2008).

16	Manchon, A. & Zhang, S. Theory of spin torque due to spin-orbit coupling. *Phys. Rev. B* **79**, 094422 (2009).

17	Garate, I. & MacDonald, A. Influence of a transport current on magnetic anisotropy in gyrotropic ferromagnets. *Phys. Rev. B* **80**, 134403 (2009).

18	Matos-Abiague, A. & Rodríguez-Suárez, R. Spin-orbit coupling mediated spin torque in a single ferromagnetic layer. *Phys. Rev. B* **80**, 094424 (2009).

19	Haney, P. M. & Stiles, M. D. Current-Induced Torques in the Presence of Spin-Orbit Coupling. *Phys. Rev. Lett.* **105**, 126602 (2010).

20	Vedyayev, A., Strelkov, N., Chshiev, M., Ryzhanova, N. & Dieny, B. Spin Transfer Torques induced by Spin Hall Effect. *http://arxiv.org/abs/1108.2589v1*, (2011).

21	Wang, X. & Manchon, A. Rashba spin torque in an ultrathin ferromagnetic metal layer. *http://arxiv.org/abs/1111.5466* (2011).

22	Wang, X. & Manchon, A. Diffusive spin dynamics in ferromagnetic thin films with a Rashba interaction. *Phys Rev Lett* **108**, 117201, (2012).

23	Manchon, A. Spin Hall effect versus Rashba torque: a Diffusive Approach. *http://arxiv.org/abs/1204.4869*, (2012).

24	Pesin, D. A. & MacDonald, A. H. Quantum kinetic theory of current-induced torques in Rashba ferromagnets. *Phys. Rev. B* **86**, 014416 (2012).

25	Haney, P. M., Lee, H.-W., Lee, K.-J., Manchon, A. & Stiles, M. D. Current induced torques and interfacial spin-orbit coupling: Semiclassical Modeling. *http://arxiv.org/abs/1301.4513v1*, (2013).

26	Kim, K.-W., Seo, S.-M., Ryu, J., Lee, K.-J. & Lee, H.-W. Magnetization dynamics induced by in-plane currents in ultrathin magnetic nanostructures with Rashba spin-orbit coupling. *Phys. Rev. B* **85**, 180404 (2012).

27	van der Bijl, E. & Duine, R. A. Current-induced torques in textured Rashba ferromagnets. *Phys. Rev. B* **86**, 094406 (2012).

28	Dyakonov, M. I. & Perel, V. I. Possibility of Orienting Electron Spins with Current. *Jetp Letters-Ussr* **13**, 467-&, (1971).

29	Bychkov, Y. A. & Rashba, E. I. Properties of a 2d Electron-Gas with Lifted Spectral Degeneracy. *JETP Lett.* **39**, 78-81, (1984).

30	Dresselhaus, G. Spin-Orbit Coupling Effects in Zinc Blende Structures. *Physical Review* **100**, 580-586, (1955).

31	Zhang, S., Levy, P. & Fert, A. Mechanisms of Spin-Polarized Current-Driven Magnetization Switching. *Phys. Rev. Lett.* **88**, 236601 (2002).

32	Tserkovnyak, Y., Brataas, A. & Bauer, G. E. W. Theory of current-driven magnetization dynamics in inhomogeneous ferromagnets. *J. Magn. Magn. Mater.* **320**, 1282-1292, (2008).

33	Liu, L., Lee, O. J., Gudmundsen, T. J., Ralph, D. C. & Buhrman, R. A. Current-Induced Switching of Perpendicularly Magnetized Magnetic Layers Using Spin Torque from the Spin Hall Effect. *Phys. Rev. Lett.* **109**, 096602 (2012).

34	Avci, C. O., Garello, K., Miron, I. M., Gaudin, G., Auffret, S., Boulle, O. & Gambardella, P. Magnetization switching of an MgO/Co/Pt layer by in-plane current injection. *Appl. Phys. Lett.* **100**, 212404 (2012).

35	Liu, L., Pai, C. F., Li, Y., Tseng, H. W., Ralph, D. C. & Buhrman, R. A. Spin-torque switching with the giant spin Hall effect of tantalum. *Science* **336**, 555-558, (2012).

36	Gaudin, G., Miron, I. M., Gambardella, P. & Schuhl, A. Magnetic memory element. WO 2012/014131.

37	Gaudin, G., Miron, I. M., Gambardella, P. & Schuhl, A. Writable magnetic element. WO 2012/014132.

38	Kim, J., Sinha, J., Hayashi, M., Yamanouchi, M., Fukami, S., Suzuki, T., Mitani, S. & Ohno, H. Layer thickness dependence of the current-induced effective field vector in Ta|CoFeB|MgO. *Nat Mater* **12**, 240-245, (2013).

39	Miron, I., Zermatten, P. J., Gaudin, G., Auffret, S., Rodmacq, B. & Schuhl, A. Domain Wall Spin Torquemeter. *Phys. Rev. Lett.* **102**, 137202 (2009).

40	Ibrahim, I. S., Schweigert, V. A. & Peeters, F. M. Diffusive transport in a Hall junction with a microinhomogeneous magnetic field. *Phys. Rev. B* **57**, 15416-15427, (1998).

41	Liu, L., Buhrman, R. A. & Ralph, D. C. Review and Analysis of Measurements of the Spin Hall Effect in Platinum. *arXiv:1111.3702v3*, (2011).

42    Niimi, Y., Wei, D., Idzuchi, H., Wakamura, T., Kato, T. & Otani, Y. Experimental Verification of Comparability between Spin-Orbit and Spin-Diffusion Lengths. *Phys. Rev. Lett.* **110**, 016805 (2013).

43    Rodmacq, B., Manchon, A., Ducruet, C., Auffret, S. & Dieny, B. Influence of thermal annealing on the perpendicular magnetic anisotropy of Pt/Co/AlOx trilayers. *Phys. Rev. B* **79**, 024423 (2009).

44    Wang, Y., Wang, W. X., Wei, H. X., Zhang, B. S., Zhan, W. S. & Han, X. F. Effect of annealing on the magnetic tunnel junction with Co/Pt perpendicular anisotropy ferromagnetic multilayers. *J. Appl. Phys.* **107**, 09c711 (2010).

45    Zimmler, M., Özyilmaz, B., Chen, W., Kent, A., Sun, J., Rooks, M. & Koch, R. Current-induced effective magnetic fields in Co∕Cu∕Co nanopillars. *Phys. Rev. B* **70**, 184438 (2004).

SUPPLEMENTARY INFORMATION

# Symmetry and magnitude of spin-orbit torques in ferromagnetic heterostructures

Kevin Garello[1], Ioan Mihai Miron[2], Can Onur Avci[1], Frank Freimuth[3], Yuriy Mokrousov[3], Stefan Blügel[3], Stéphane Auffret[2], Olivier Boulle[2], Gilles Gaudin[2], and Pietro Gambardella[1,4,5]

[1] *Catalan Institute of Nanotechnology (ICN), E-08193 Barcelona, Spain*
[2] *SPINTEC, UMR-8191, CEA/CNRS/UJF/GINP, INAC, F-38054 Grenoble, France*
[3] *Peter Grünberg Institut and Institute for Advanced Simulation, Forschungszentrum Jülich and JARA, 52425 Jülich, Germany*
[4] *Institució Catalana de Recerca i Estudis Avançats (ICREA), E-08010 Barcelona, Spain*
[5] *Department of Materials, ETH Zurich, CH-8093 Zurich, Switzerland*

**Table of Contents:**

S1. General expression for the spin accumulation and SOTs

S2. Harmonic analysis of the Hall voltage

S3. Separation of AHE and PHE

S4. Modulation of the Hall voltage by an external ac field

S5. Measurement of current-induced effective fields in the case of nonzero PHE

S6. Macrospin simulations

S7. Sample-dependent offset and Nernst-Ettingshausen effect

S8. Current dispersion in the Hall voltage probes

S9. Measurements in the case of nonuniform magnetization

S10. Comparison of AC and DC detection methods

S11. Dynamic simulations of the $m_y$ component generated by $T^\parallel$

S12. Influence of thermal effects

S13. Supplementary references

## S1. General expression for the spin accumulation and SOTs

In this section we derive general expressions for the spin accumulation ($\delta\boldsymbol{m}$), effective fields ($\boldsymbol{B}^I$), and torques ($\boldsymbol{T}$) induced by an electric current in trilayers with structure inversion asymmetry along the stacking direction *z*. The current is driven by an applied electric field ($\boldsymbol{E}$) in the *xy* plane. We consider only the case of trilayers that exhibit continuous rotational symmetry about the *z* axis and mirror symmetry for all the planes that are perpendicular to the *xy* plane, i.e., contain the *z* axis. The results of this section apply to samples that are either polycrystalline, as in our case, or disordered. In fully epitaxial systems that display discrete rotational symmetry around the stacking direction it is expected also that $\delta\boldsymbol{m}$, $\boldsymbol{B}^I$, and $\boldsymbol{T}$ vary as the sample is rotated while keeping the directions of the electric field and magnetization fixed in space.

Consider an applied current that leads to a spin accumulation $\delta\boldsymbol{m}$. Explicitly, in the equations below, $\delta\boldsymbol{m}$ denotes the induced magnetization associated with the spin accumulation. $\delta\boldsymbol{m}$ induces a change of the exchange field ($\delta\boldsymbol{B}^{xc}$) in the ferromagnet, which acts as an effective magnetic field on the magnetization ($\boldsymbol{m}$), given by $\boldsymbol{B}^I = \delta\boldsymbol{B}^{xc} = \frac{B^{xc}}{m}\delta\boldsymbol{m}$. The resulting torque is given by $\boldsymbol{T} = \boldsymbol{m} \times \boldsymbol{B}^I$. Since the torque depends only on the component of $\boldsymbol{B}^I$ that is perpendicular to $\boldsymbol{m}$, in the following $\boldsymbol{B}^I$ and $\delta\boldsymbol{m}$ will denote the perpendicular components of the effective field and the spin accumulation, respectively.

In Fig. S1a and b we consider the case of magnetization in the *xz* plane and electric field in the *x* direction. We show that symmetry allows for two components for the spin accumulation: A longitudinal one, $\delta\boldsymbol{m}^\parallel$, which lies in the *xz* plane, and a perpendicular one, $\delta\boldsymbol{m}^\perp$, which points in the *y* direction. $\boldsymbol{E}$ is invariant under mirror reflection at the *xz* plane. However, $\boldsymbol{m}$ is inverted, because it is an axial vector. Similarly, the component $\delta\boldsymbol{m}^\parallel$ of the axial vector $\delta\boldsymbol{m}$ is inverted, but $\delta\boldsymbol{m}^\perp$ is invariant. Thus, $\delta\boldsymbol{m}^\parallel$ has to be an odd function of $\boldsymbol{m}$, while $\delta\boldsymbol{m}^\perp$ has to be an even function. Mirror reflection at the *yz* plane followed by a rotation around the *z* axis by 180º leads to the same conclusion. There is no symmetry operation that forbids either $\delta\boldsymbol{m}^\parallel$ or $\delta\boldsymbol{m}^\perp$, owing to the structure inversion asymmetry. For example, if there was inversion symmetry, $\boldsymbol{E}$ would change under inversion while $\boldsymbol{m}$ and $\delta\boldsymbol{m}$ would remain unchanged. Thus, inversion symmetry would dictate that both $\boldsymbol{E}$ and $-\boldsymbol{E}$ lead to the same spin accumulation, meaning that, in such a case, the linear response of the spin accumulation would have to be zero.

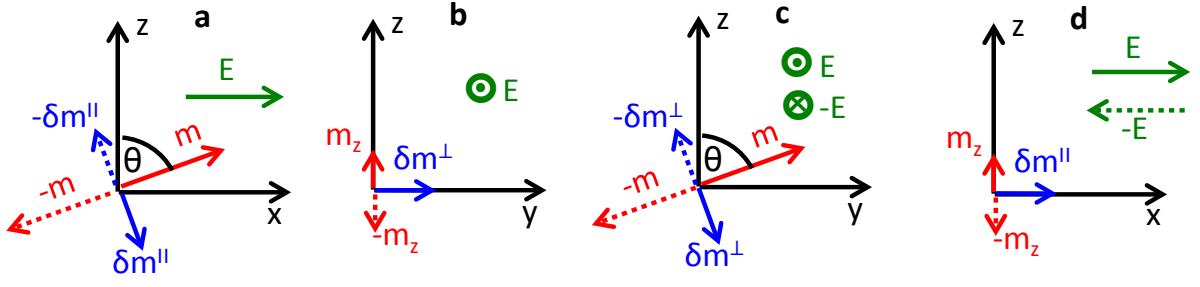

**Figure S1.** Transformation of electric field **E**, magnetization **m** and spin accumulation $\delta\boldsymbol{m}$ under mirror reflections. a) Magnetization in the *xz* plane. Mirror reflection at the *xz* plane leaves **E** invariant, but inverts **m** and $\delta\boldsymbol{m}^{\|}$, because the latter two transform like axial vectors. b) Same as a, but view on the *yz* plane. $\delta\boldsymbol{m}^{\perp}$ is invariant under reflection at the *xz* plane. c) Magnetization in the *yz* plane. Mirror reflection at the *yz* plane inverts **E**, **m** and $\delta\boldsymbol{m}^{\perp}$. d) Same as c, but view on the *xz* plane. The component $\delta\boldsymbol{m}^{\|}$ is invariant under reflection at the *yz* plane.

In Fig. S1c and d we consider the case of magnetization in the *yz* plane and the electric field again along the *x* direction. The longitudinal spin accumulation, $\delta\boldsymbol{m}^{\|}$, points now in the *x* direction, while the transverse spin accumulation, $\delta\boldsymbol{m}^{\perp}$, lies in the *yz* plane. Mirror reflection at the *yz* plane inverts **E**, **m**, and $\delta\boldsymbol{m}^{\perp}$, while $\delta\boldsymbol{m}^{\|}$ is invariant. Within linear response, $\delta\boldsymbol{m}$ must change sign upon inversion of **E**. It follows that $\delta\boldsymbol{m}^{\perp}$ is again an even function of **m** because, for inverted electric field and inverted magnetization, $\delta\boldsymbol{m}^{\perp}$ is inverted. Likewise, $\delta\boldsymbol{m}^{\|}$ is again an odd function of **m**. Mirror reflection at the *xz* plane followed by a rotation around the *z* axis by 180° leads to the same conclusion. In the special case of **m** ∥ *y*, $\delta\boldsymbol{m}$ vanishes because in this case mirror reflection at the *yz* plane followed by a rotation around *z* by 180° leaves **m** and **E** invariant, but $\delta\boldsymbol{m}$ is inverted. However, if **m** has nonzero out-of-plane components, there is no symmetry operation present which forbids any of the two components $\delta\boldsymbol{m}^{\|}$ or $\delta\boldsymbol{m}^{\perp}$. We will show below that, for **m** lying in the *xz* plane, one has

$$\delta\boldsymbol{m}^{\|} = \frac{E}{B^{xc}}(\boldsymbol{e}_y \times \boldsymbol{m})\left(A_0^{\varphi} + A_2^{\varphi} \sin^2\theta + A_4^{\varphi} \sin^4\theta + \cdots\right) \qquad (1)$$

and

$$\delta\boldsymbol{m}^{\perp} = -\frac{E}{B^{xc}}\boldsymbol{e}_y\left(A_0^{\theta} + A_2^{\theta} \sin^2\theta + A_4^{\theta} \sin^4\theta + \cdots\right) \qquad (2)$$

For **m** lying in the *yz* plane one has instead:

$$\delta\boldsymbol{m}^{\|} = \frac{E}{B^{xc}}(\boldsymbol{e}_y \times \boldsymbol{m})\left(A_0^{\varphi} + B_2^{\theta} \sin^2\theta + B_4^{\theta} \sin^4\theta + \cdots\right) \qquad (3)$$

and

$$\delta\boldsymbol{m}^{\perp} = -\frac{E}{B^{xc}}(\boldsymbol{e}_y \times \boldsymbol{m}) \times \boldsymbol{m}\left(A_0^{\theta} - B_2^{\varphi} \sin^2\theta - B_4^{\varphi} \sin^4\theta - \cdots\right). \qquad (4)$$

The coefficients $A_2^\varphi, A_4^\varphi, A_2^\theta, A_4^\theta$ and $B_2^\varphi, B_4^\varphi, B_2^\theta, B_4^\theta$ describe what we refer to as anisotropy of the SOT. In the absence of anisotropy, one single parameter, $A_0^\varphi$, governs the longitudinal accumulation $\delta\boldsymbol{m}^\parallel$, whereas one single parameter, $A_0^\theta$, governs the transverse accumulation $\delta\boldsymbol{m}^\perp$. To describe the anisotropy one needs four parameters for each order of the expansion, where two *A* parameters describe the anisotropy of the two spin-accumulation components for the case of magnetization in the *xz* plane and two *B* parameters describe the anisotropy for the case of magnetization in the *yz* plane. Since the trilayers considered in this work exhibit continuous rotational symmetry around the *z* axis, no additional anisotropy arises from the angle $\varphi$ of the magnetization.

Whereas symmetry arguments provide general expressions of $\delta\boldsymbol{m}^\parallel$ and $\delta\boldsymbol{m}^\perp$, as will be shown below, the same arguments do not provide information about the magnitude of these two terms and the underlying mechanisms. Experiment and theory suggest that both components can be important, and their origin is a matter of debate. As the angle $\theta$ in Fig. S1 is varied, the magnitudes of $\delta\boldsymbol{m}^\parallel$ and $\delta\boldsymbol{m}^\perp$ are generally expected to change. The resulting anisotropy in the spin accumulation can arise from spin-dependent interface resistances that influence the spin current from the spin Hall effect as it traverses the interface, or from anisotropic relaxation times, as pointed out in Ref. 1. From the electronic structure point of view, the band energies and thus also the Fermi surface change as a function of $\theta$ due to spin-orbit coupling. This can be understood, e.g., in an *sd* model, where only the *s* electrons are subject to Rashba type spin orbit interaction: In the absence of hybridization the spin of the *s*-bands is determined by the Rashba interaction, while the spin of the *d*-bands is determined by the direction of the exchange field. Due to hybridization, the coupled *sd* model exhibits a band structure which depends on $\theta$. The symmetries of the trilayers restrict the allowed $\theta$ dependence.

In order to obtain an expression for the spin accumulation valid for any orientation of $\boldsymbol{m}$, one can decompose the electric field into two components, one parallel to the plane spanned by $\boldsymbol{m}$ and the *z* axis (Eqs. 1 and 2) and one perpendicular to it (Eqs. 3 and 4). Since we are considering the linear response of the spin accumulation to an electric field, $\delta\boldsymbol{m}$ is given by a superposition of these two simple cases in general. In the following, we work out a general expression for the linear response of axial vectors (such as $\delta\boldsymbol{m}$, $\boldsymbol{B}^I$, and $\boldsymbol{T}$) to an applied electric field which conforms with the symmetry requirements discussed above. The following derivation holds for all axial vectors perpendicular to the magnetization; however, to be concrete, we refer to torque $\boldsymbol{T}$ and torkance $\boldsymbol{t}$. Readers who are more familiar with spin accumulation may consider $\boldsymbol{T}$ as spin-accumulation and $\boldsymbol{t}$ as spin-accumulation per applied electric field.

If the electric field is along the $x$ axis, the torque acting on the magnetization can be written as $\boldsymbol{T}(\theta, \varphi) = \boldsymbol{t}(\theta, \varphi) E$, where $\boldsymbol{t}(\theta, \varphi)$ is the torkance and $(\theta, \varphi)$ are the polar and azimuthal coordinates of the unit magnetization vector $\boldsymbol{m} = (\sin\theta \cos\varphi, \sin\theta \sin\varphi, \cos\theta)$. Since the torkance is perpendicular to the magnetization, it may be expressed in terms of the basis vectors of the spherical coordinate system $\boldsymbol{e}_\theta = (\cos\theta \cos\varphi, \cos\theta \sin\varphi, -\sin\theta)$ and $\boldsymbol{e}_\varphi = (-\sin\varphi, \cos\varphi, 0)$ with components $t_\theta(\theta, \varphi)$ and $t_\varphi(\theta, \varphi)$:

$$\boldsymbol{t}(\theta, \varphi) = t_\theta(\theta, \varphi) \boldsymbol{e}_\theta + t_\varphi(\theta, \varphi) \boldsymbol{e}_\varphi = \sum_{s=\theta,\varphi} t_s(\theta, \varphi) \boldsymbol{e}_s . \tag{5}$$

It is assumed that the magnetic layer exhibits continuous rotational symmetry around the $z$ axis and mirror symmetry with respect to planes perpendicular to the layer plane. We consider first the consequences of rotational symmetry. If the electric field makes an angle $\gamma$ with the $x$ axis (Fig. S2a), we have $\boldsymbol{E} = (\cos\gamma, \sin\gamma, 0)E$ and the torque can be rewritten as $\boldsymbol{T}(\theta, \varphi) = \sum_{s=\theta,\varphi} t_s(\theta, \varphi - \gamma) E \boldsymbol{e}_s$. Since the torque is linear in the electric field, we can decompose $\boldsymbol{T}(\theta, \varphi)$ into one component due to the electric field $E\cos\gamma$ along the $x$ direction plus a second component due to the electric field $E\sin\gamma$ along the $y$ direction: $\boldsymbol{T}(\theta, \varphi) = \sum_{s=\theta,\varphi} t_s(\theta, \varphi) E \cos\gamma \, \boldsymbol{e}_s + \sum_{s=\theta,\varphi} t_s(\theta, \varphi - \pi/2) E \sin\gamma \, \boldsymbol{e}_s$. The continuous rotation axis thus leads to the condition $t_s(\theta, \varphi - \gamma) = t_s(\theta, \varphi)\cos\gamma + t_s(\theta, \varphi - \pi/2)\sin\gamma$, which restricts the allowed $\varphi$-dependence such that

$$t_s(\theta, \varphi) = F_1^s(\theta) \cos\varphi + F_2^s(\theta) \sin\varphi, \tag{6}$$

where $F_1^s$ and $F_2^s$ are two functions of $\theta$.

As the magnetization and torque are axial vectors, the $x$ and $z$ components change sign under mirror reflection with respect to the $xz$ plane, while the $y$ component is conserved (Fig. S2b). Mirror reflection symmetry thus dictates that $t_x(\pi - \theta, \pi - \varphi) = -t_x(\theta, \varphi)$, $t_y(\pi - \theta, \pi - \varphi) = t_y(\theta, \varphi)$ and $t_z(\pi - \theta, \pi - \varphi) = -t_z(\theta, \varphi)$, which is equivalent to

$$t_s(\pi - \theta, \pi - \varphi) = -t_s(\theta, \varphi) . \tag{7}$$

Mirror reflection with respect to the $yz$ plane changes the sign of the $y$ and $z$ components of the magnetization and torkance as well as the sign of the electric field along the $x$ direction (Fig. S2c), leading to the conditions $t_x(\pi - \theta, -\varphi) = -t_x(\theta, \varphi)$, $t_y(\pi - \theta, -\varphi) = t_y(\theta, \varphi)$ and $t_z(\pi - \theta, -\varphi) = t_z(\theta, \varphi)$, or, equivalently, to

$$t_s(\pi - \theta, -\varphi) = t_s(\theta, \varphi) . \tag{8}$$

Since the angles $(-\theta, \varphi)$ and $(\theta, \pi + \varphi)$ are equivalent, an additional condition is given by

$$t_s(-\theta, \varphi) = -t_s(\theta, \pi + \varphi), \tag{9}$$

where the minus sign on the right hand side compensates for the minus sign on the right hand side in $\boldsymbol{e}_s(-\theta, \varphi) = -\boldsymbol{e}_s(\theta, \pi + \varphi)$.

We expand the functions $F_j^s$ in Eq. 6 as a Fourier series,

$$F_j^s(\theta) = A_{j,0}^s + \sum_{n=1}^{\infty}\left(A_{j,n}^s \cos n\theta + B_{j,n}^s \sin n\theta\right), \tag{10}$$

and note that Eq. 9 is satisfied if $F_j^s(-\theta) = F_j^s(\theta)$, and Eq. 7 as well as Eq. 8 are satisfied if $F_1^s(\pi - \theta) = F_1^s(\theta)$ and $F_2^s(\pi - \theta) = -F_2^s(\theta)$. The condition $F_j^s(-\theta) = F_j^s(\theta)$ rules out terms proportional to $\sin n\theta$ in the Fourier expansion of $F_j^s(\theta)$. Additionally, $\cos n\theta$ terms with odd integers $n$ are ruled out in the expansion of $F_1^s$ by the condition $F_1^s(\pi - \theta) = F_1^s(\theta)$. For even $n$, $\cos n\theta$ can be written as a sum of products of even powers of $\sin \theta$ and $\cos \theta$ due to the relation $\cos n\theta = \cos^n \theta - \frac{n(n-1)}{2} \sin^2 \theta \cos^{n-2} \theta + \cdots$. Because $\cos^2 \theta = 1 - \sin^2 \theta$, even powers of $\cos \theta$ can be expressed as sums of even powers of $\sin \theta$. Consequently, $F_1^s(\theta)$ is a sum of even powers of $\sin \theta$. The condition $F_2^s(\pi - \theta) = -F_2^s(\theta)$ rules out $\cos n\theta$ with even $n$. Due to the relations discussed above, it is clear that $\cos n\theta$ with odd $n$ can be written as $\cos \theta$ times a sum of even powers of $\sin \theta$. Hence, we arrive at the expansion

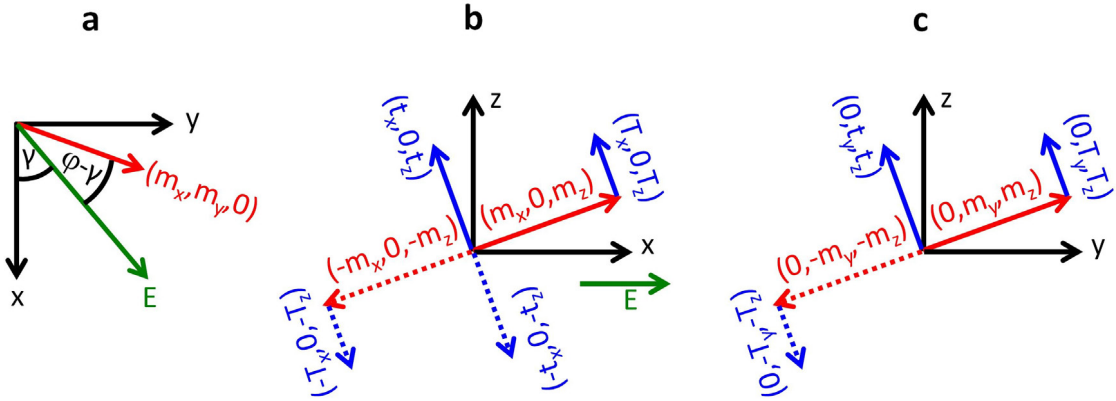

**Figure S2.** Symmetry properties of the torque. **a,** Angle definitions used in the text. **b,** Effect of mirror reflection at the *xz* plane on the *x* and *z* components of *T* and *m*. Since the electric field *E* is invariant, $t_x$ and $t_z$ undergo the same sign change as $T_x$ and $T_z$. **c,** Effect of mirror reflection at the *yz* plane on the *y* and *z* components of *T* and *m*. Since *E* (along *x*) changes sign, $t_y$ and $t_z$ are invariant.

$$t_s(\theta, \varphi) = \cos \varphi \, (A_0^s + A_2^s \sin^2 \theta + A_4^s \sin^4 \theta + \cdots)$$
$$+ \cos \theta \sin \varphi \, (B_0^s + B_2^s \sin^2 \theta + B_4^s \sin^4 \theta + \cdots). \tag{11}$$

The torkance $\mathbf{t}(\theta, \varphi)$ is the sum of the even and odd parts $\mathbf{t}^\parallel(\theta, \varphi) = \frac{t(\theta,\varphi)+t(\pi-\theta,\pi+\varphi)}{2}$ and $\mathbf{t}^\perp(\theta, \varphi) = \frac{t(\theta,\varphi)-t(\pi-\theta,\pi+\varphi)}{2}$, respectively. Due to $\mathbf{e}_\varphi(\pi - \theta, \pi + \varphi) = -\mathbf{e}_\varphi(\theta, \varphi)$ and $\mathbf{e}_\theta(\pi - \theta, \pi + \varphi) = \mathbf{e}_\theta(\theta, \varphi)$, the expansions of the even and odd parts of the torque are given by

$$\mathbf{t}^\parallel(\theta, \varphi) = \cos \varphi \, \left(A_0^\varphi + A_2^\varphi \sin^2 \theta + A_4^\varphi \sin^4 \theta + \cdots\right) \mathbf{e}_\varphi +$$
$$\cos \theta \sin \varphi \, \left(B_0^\theta + B_2^\theta \sin^2 \theta + B_4^\theta \sin^4 \theta + \cdots\right) \mathbf{e}_\theta \tag{12}$$

and

$$\mathbf{t}^\perp(\theta, \varphi) = \cos \varphi \, \left(A_0^\theta + A_2^\theta \sin^2 \theta + A_4^\theta \sin^4 \theta + \cdots\right) \mathbf{e}_\theta +$$
$$\cos \theta \sin \varphi \, \left(B_0^\varphi + B_2^\varphi \sin^2 \theta + B_4^\varphi \sin^4 \theta + \cdots\right) \mathbf{e}_\varphi \tag{13}$$

An additional requirement is that the torkance shall be independent of $\varphi$ at $\theta=0$. Since $\cos \varphi \, \mathbf{e}_\theta(0, \varphi) - \sin \varphi \, \mathbf{e}_\varphi(0, \varphi) = \mathbf{e}_x$ and $\sin \varphi \, \mathbf{e}_\theta(0, \varphi) + \cos \varphi \, \mathbf{e}_\varphi(0, \varphi) = \mathbf{e}_y$, this is achieved by imposing $A_0^\varphi = B_0^\theta$ and $A_0^\theta = -B_0^\varphi$, which leads to

$$\mathbf{t}^\parallel(\theta, \varphi) = \cos \varphi \, \left(A_0^\varphi + A_2^\varphi \sin^2 \theta + A_4^\varphi \sin^4 \theta + \cdots\right) \mathbf{e}_\varphi +$$
$$\cos \theta \sin \varphi \, \left(A_0^\varphi + B_2^\theta \sin^2 \theta + B_4^\theta \sin^4 \theta + \cdots\right) \mathbf{e}_\theta \tag{14}$$

and

$$\mathbf{t}^\perp(\theta, \varphi) = \cos \varphi \, \left(A_0^\theta + A_2^\theta \sin^2 \theta + A_4^\theta \sin^4 \theta + \cdots\right) \mathbf{e}_\theta +$$
$$\cos \theta \sin \varphi \, \left(-A_0^\theta + B_2^\varphi \sin^2 \theta + B_4^\varphi \sin^4 \theta + \cdots\right) \mathbf{e}_\varphi \tag{15}$$

It is straightforward to verify that Eqs. 14 and 15 lead to the following expression for the torques:

$$\mathbf{T}^\parallel = \mathbf{m} \times [(\mathbf{e}_z \times \mathbf{E}) \times \mathbf{m}] \left[A_0^\varphi + B_2^\theta (\mathbf{e}_z \times \mathbf{m})^2 + B_4^\theta (\mathbf{e}_z \times \mathbf{m})^4 + \ldots \right] +$$
$$(\mathbf{m} \times \mathbf{e}_z)(\mathbf{m} \cdot \mathbf{E}) \left[\left(B_2^\theta - A_2^\varphi\right) + \left(B_4^\theta - A_4^\varphi\right)(\mathbf{e}_z \times \mathbf{m})^2 + \ldots \right] \tag{16}$$

and

$$\mathbf{T}^\perp = (\mathbf{e}_z \times \mathbf{E}) \times \mathbf{m} \left[A_0^\theta - B_2^\varphi (\mathbf{e}_z \times \mathbf{m})^2 - B_4^\varphi (\mathbf{e}_z \times \mathbf{m})^4 - \ldots \right] +$$
$$\mathbf{m} \times [(\mathbf{m} \times \mathbf{e}_z)(\mathbf{m} \cdot \mathbf{E})] \left[\left(A_2^\theta + B_2^\varphi\right) + \left(A_4^\theta + B_4^\varphi\right)(\mathbf{e}_z \times \mathbf{m})^2 + \ldots \right]. \tag{17}$$

Notably, $T^\parallel$ and $T^\perp$ depend explicitly on the three vectors $E$, $m$, and $e_z$, where the presence of $e_z$ reflects the absence of $z$ reflection symmetry. By choosing $E$ (the electric current) to point in the $x$ direction and $|E| = 1$ for simplicity, Eqs. 16 and 17 give

$$T^\parallel = m \times (e_y \times m)[A_0^\varphi + B_2^\theta (e_z \times m)^2 + B_4^\theta (e_z \times m)^4 + \ldots] +$$
$$(m \times e_z)(m \cdot e_x)[(B_2^\theta - A_2^\varphi) + (B_4^\theta - A_4^\varphi)(e_z \times m)^2 + \ldots] \quad (18)$$

and

$$T^\perp = e_y \times m \left[A_0^\theta - B_2^\varphi (e_z \times m)^2 - B_4^\varphi (e_z \times m)^4 - \ldots\right] +$$
$$m \times [(m \times e_z)(m \cdot e_x)][(A_2^\theta + B_2^\varphi) + (A_4^\theta + B_4^\varphi)(e_z \times m)^2 + \ldots]. \quad (19)$$

The previous two equations have been used in the main part of the manuscript to give a vector form of the torques. Only the coefficients $A_0^\varphi \equiv T_0^\parallel$, $A_2^\varphi \equiv T_2^\parallel$, $A_4^\varphi \equiv T_4^\parallel$ and $A_0^\theta \equiv T_0^\perp$, $B_2^\varphi \equiv -T_2^\perp$, $B_4^\varphi \equiv -T_4^\perp$, have been retained there, because the others are below the experimental detection limit.

We note that the special case of $A_n^\varphi = B_n^\theta = 0$ for all $n \neq 0$ leads to

$$T^\parallel = A_0^\varphi (\cos\varphi\, e_\varphi + \cos\theta \sin\varphi\, e_\theta) = A_0^\varphi [(e_y \cdot e_\varphi) e_\varphi + (e_y \cdot e_\theta) e_\theta]$$
$$= A_0^\varphi [e_y - (e_y \cdot m)m] = A_0^\varphi m \times (e_y \times m) \quad (20)$$

and, likewise, the special case of $A_n^\theta = B_n^\varphi = 0$ for all $n \neq 0$ simplifies $T^\perp$ to

$$T^\perp = A_0^\theta (\cos\varphi\, e_\theta - \cos\theta \sin\varphi\, e_\varphi) = A_0^\theta (\cos\varphi\, e_\varphi + \cos\theta \sin\varphi\, e_\theta) \times m$$
$$= A_0^\theta [m \times (e_y \times m)] \times m = A_0^\theta (e_y \times m). \quad (21)$$

These reduced expressions have been obtained theoretically for several models discussed in the literature (see main text).

For the purpose of comparison with the experiment, it is useful to derive the effective magnetic fields $B^I(\theta, \varphi)$ associated with the torques. Since $T = m \times B^I$, by multiplying the previous equation by $m$ and noting that $(m \times B^I) \times m = B^I(m \cdot m) - m(m \cdot B^I) = B^I$, one has $B^I = T \times m$. The effective fields corresponding to $T^\parallel$ and $T^\perp$ are

$$B^\parallel = (e_y \times m)[A_0^\varphi + B_2^\theta (e_z \times m)^2 + B_4^\theta (e_z \times m)^4 + \ldots] +$$
$$(m \times e_z) \times m(m \cdot e_x)[(B_2^\theta - A_2^\varphi) + (B_4^\theta - A_4^\varphi)(e_z \times m)^2 + \ldots] \quad (22)$$

and

$$\boldsymbol{B}^\perp = (\boldsymbol{e}_y \times \boldsymbol{m}) \times \boldsymbol{m}\left[A_0^\theta - B_2^\varphi (\boldsymbol{e}_z \times \boldsymbol{m})^2 - B_4^\varphi (\boldsymbol{e}_z \times \boldsymbol{m})^4 - \dots\right] +$$

$$(\boldsymbol{m} \times \boldsymbol{e}_z)(\boldsymbol{m} \cdot \boldsymbol{e}_x)\left[(A_2^\theta + B_2^\varphi) + (A_4^\theta + B_4^\varphi)(\boldsymbol{e}_z \times \boldsymbol{m})^2 + \dots\right]. \quad (23)$$

These expressions are valid also for the spin accumulation, since $\delta \boldsymbol{m}^\| \sim \boldsymbol{B}^\|$ and $\delta \boldsymbol{m}^\perp \sim \boldsymbol{B}^\perp$. In spherical coordinates, using $\boldsymbol{e}_\theta \times \boldsymbol{m} = -\boldsymbol{e}_\varphi$ and $\boldsymbol{e}_\varphi \times \boldsymbol{m} = \boldsymbol{e}_\theta$, we obtain

$$\boldsymbol{B}^\|(\theta, \varphi) = \cos \varphi \left(A_0^\varphi + A_2^\varphi \sin^2 \theta + A_4^\varphi \sin^4 \theta + \cdots\right)\boldsymbol{e}_\theta -$$

$$\cos \theta \sin \varphi \left(A_0^\varphi + B_2^\theta \sin^2 \theta + B_4^\theta \sin^4 \theta + \cdots\right)\boldsymbol{e}_\varphi \quad (24)$$

and

$$\boldsymbol{B}^\perp(\theta, \varphi) = -\cos \varphi \left(A_0^\theta + A_2^\theta \sin^2 \theta + A_4^\theta \sin^4 \theta + \cdots\right)\boldsymbol{e}_\varphi +$$

$$\cos \theta \sin \varphi \left(-A_0^\theta + B_2^\varphi \sin^2 \theta + B_4^\varphi \sin^4 \theta + \cdots\right)\boldsymbol{e}_\theta. \quad (25)$$

The polar and azimuthal components of the total effective field $\boldsymbol{B}^I = \boldsymbol{B}^\|(\theta, \varphi) + \boldsymbol{B}^\perp(\theta, \varphi)$ are then given by

$$B_\theta(\theta, \varphi) = \cos \varphi \left(A_0^\varphi + A_2^\varphi \sin^2 \theta + A_4^\varphi \sin^4 \theta + \cdots\right) +$$

$$\cos \theta \sin \varphi \left(-A_0^\theta + B_2^\varphi \sin^2 \theta + B_4^\varphi \sin^4 \theta + \cdots\right) \quad (26)$$

and

$$B_\varphi(\theta, \varphi) = -\cos \varphi \left(A_0^\theta + A_2^\theta \sin^2 \theta + A_4^\theta \sin^4 \theta + \cdots\right) -$$

$$\cos \theta \sin \varphi \left(A_0^\varphi + B_2^\theta \sin^2 \theta + B_4^\theta \sin^4 \theta + \cdots\right). \quad (27)$$

We conclude this section by noting that Eqs. 22, 23 determine only the components of the effective field which are perpendicular to $\boldsymbol{m}$. In general, however, even though effective fields along the magnetization direction do not produce any torques, it is expected that the current also induces fields parallel to $\boldsymbol{m}$. While such a radial component of the current-induced magnetic field ($\boldsymbol{B}_r$) is not accessible in the present experiment, we provide below its most general expression compatible with the symmetry of the system. By extending the analysis for the torque presented above to the effective field, one can show that

$$\boldsymbol{B}_r(\theta, \varphi) = [\sin \theta \cos \theta \cos \varphi \left(A_0^r + A_2^r \sin^2 \theta + A_4^r \sin^4 \theta + \cdots\right) +$$

$$\sin \theta \sin \varphi (B_0^r + B_2^r \sin^2 \theta + B_4^r \sin^4 \theta + \cdots)]\boldsymbol{m}. \quad (28)$$

## S2. Harmonic analysis of the Hall voltage

We use the Hall voltage ($V_H$) to measure the behaviour of the magnetization as a function of external field and current-induced torques. In general, one has

$$V_H = IR_H = R_{AHE}I\cos\theta + R_{PHE}I\sin^2\theta\sin 2\varphi, \qquad (29)$$

where $I$ is the current and $R_H$ the Hall resistance due to the anomalous Hall effect (AHE) and planar Hall effect (PHE). We omit here the ordinary Hall effect, which is negligible in ferromagnetic materials, as well as thermoelectric effects, which will be discussed in Section S7. The AHE is proportional to $R_{AHE}I\cos\theta$ and the PHE to $R_{PHE}I\sin^2\theta\sin 2\varphi$, where $R_{AHE}$ and $R_{PHE}$ are the AHE and PHE resistances, $\theta$ and $\varphi$ the polar and azimuthal angle of the magnetization, respectively, as defined in Fig.1b of the main text. Due to the presence of effective fields, the injection of a moderate ac current $I_{ac} = Ie^{i2\pi ft}$ induces small oscillations of the magnetization around its equilbrium position $(\theta_0, \varphi_0)$, defined by the anisotropy field $B_k$ and external field $B_{ext}$. These oscillations modulate $R_H$ giving rise to a time-dependent Hall voltage signal. After simplification of the time-dependent phase factor, the dependence of the Hall voltage on the current can be expanded to first order as

$$V_H(I) \approx V_H(\theta_0, \varphi_0) + I \left.\frac{dV_H}{dI}\right|_{\theta_0,\varphi_0}. \qquad (30)$$

Straightforward differentiation of Eq. 29, keeping into account that both $\theta$ and $\varphi$ depend on $I$, gives $\frac{dV_H}{dI} = R_H^f + R_H^{2f}(I)$, where the first and second harmonic Hall resistance components are given by

$$R_H^f = R_{AHE}\cos\theta_0 + R_{PHE}\sin^2\theta_0\sin 2\varphi_0 \qquad (31)$$

and

$$R_H^{2f} = I(R_{AHE} - 2R_{PHE}\cos\theta_0\sin 2\varphi_0)\left.\frac{d\cos\theta}{dI}\right|_{\theta_0} + IR_{PHE}\sin^2\theta_0\left.\frac{d\sin 2\varphi}{dI}\right|_{\varphi_0}. \qquad (32)$$

We notice that $R_H^f$ is equivalent to the Hall resistance measured in conventional dc measurements, whereas $R_H^{2f}$ contains two terms that depend explicitly on the current. This dependence can be expressed in terms of the current-induced effective field $\boldsymbol{B}^I = \boldsymbol{B}^\| + \boldsymbol{B}^\perp + \boldsymbol{B}^{Oersted}$ by noting that

$$\frac{d\cos\theta}{dI} = \frac{d\cos\theta}{d\boldsymbol{B}^I} \cdot \frac{d\boldsymbol{B}^I}{dI} = \frac{d\cos\theta}{dB^I_\theta} b_\theta \qquad (33)$$

and

$$\frac{d\sin 2\varphi}{dI} = \frac{d\sin 2\varphi}{d\boldsymbol{B}^I} \cdot \frac{d\boldsymbol{B}^I}{dI} = \frac{d\sin 2\varphi}{dB^I_\varphi} b_\varphi, \qquad (34)$$

where $B^I_\theta$ and $B^I_\varphi$ indicate the polar and azimuthal components of $\boldsymbol{B}^I$ and $b_\theta$ and $b_\varphi$ their derivative with respect to the current. The radial component of $\boldsymbol{B}^I$ cannot affect the motion of the magnetization and is thus irrelevant to the discussion of the torques. To measure quantitatively the effective fields $b_\theta$ and $b_\varphi$ by means of Eq. 32 we need first to calculate the derivatives of $\cos\theta$ and $\sin 2\varphi$ that appear in Eqs. 33 and 34. As the magnetic field dependence of $\cos\theta$ and $\sin 2\varphi$ (proportional to $m_z$ and $m_x m_y$, respectively) is independent on the nature of the field, we can replace $\boldsymbol{B}^I$ by $\boldsymbol{B}_{ext}$ in the derivatives and obtain

$$\frac{d\cos\theta}{dI} = \frac{d\cos\theta}{dB_{ext}} \frac{1}{\sin(\theta_B - \theta_0)} b_\theta, \qquad (35)$$

$$\frac{d\sin 2\varphi}{dI} = \frac{d\sin 2\varphi}{dB^I_\varphi} b_\varphi = 2\cos 2\varphi \frac{d\varphi}{dB^I_\varphi} b_\varphi \approx \frac{2\cos 2\varphi}{B_{ext}\sin\theta_B} b_\varphi, \qquad (36)$$

where $\theta_B$ is the polar angle defining the direction of $\boldsymbol{B}_{ext}$. Note that the last relation is exact in the case of uniaxial magnetic anisotropy. Using these expressions, $R^{2f}_H$ can be written as

$$R^{2f}_H = I(R_{AHE} - 2R_{PHE}\cos\theta_0 \sin 2\varphi_0) \left.\frac{d\cos\theta}{dB_{ext}}\right|_{\theta_0} \frac{1}{\sin(\theta_B - \theta_0)} b_\theta$$
$$+ IR_{PHE}\sin^2\theta_0 \frac{2\cos 2\varphi_0}{B_{ext}\sin\theta_B} b_\varphi. \qquad (37)$$

Equation 34 is valid for any arbitrary field $\boldsymbol{B}^I$, $\boldsymbol{B}_{ext}$, orientation of $\boldsymbol{m}$, and uniaxial magnetic anisotropy, which makes it very useful in the investigation of SOTs.

### S3. Separation of AHE and PHE

To make correct use of Eq. 37, it is desirable to measure $R_{PHE}$ in order to separate the PHE and AHE contributions to the total Hall resistance. This can be readily achieved by measuring $R^f_H$ as a function of the external field applied at angles $\varphi \neq 0°, 90°$. Figure S3a shows $R^f_H$ measured at $\theta_B = 80°$, $\varphi = 50°$ by means of example. Due to the PHE, which is even with respect to magnetization reversal, the endpoints of the hysteresis loop are asymmetric. As the AHE is odd with respect to magnetization reversal, the first harmonic contributions of the AHE and PHE, $R_f^{AHE}$ and $R_f^{PHE}$, can be separated by antisymmetrization and symmetrization of $R_H^f$, respectively. This is achieved in practice by inverting $R_H^f(-B_{ext} \rightarrow +B_{ext})$ with respect to the origin of the curve and taking the sum or difference with $R_H^f(+B_{ext} \rightarrow -B_{ext})$, as shown in Fig. S3b and c, respectively. Macrospin simulations of the AHE and PHE further demonstrate that this procedure yields correct quantitative results, as shown in Fig. S3d. Finally, the PHE saturation resistance is deduced from the linear fit of $R_f^{PHE}$ vs. $\sin^2\theta_M$ (Fig. S3e). For this sample, we obtain $R_{PHE} = 0.11\ \Omega$. The saturation value of the AHE resistance is $R_{AHE} = 0.81\ \Omega$. Finally, $R_f^{AHE}$ is employed to calculate the equilibrium angle of the magnetization ($\theta_0$) at each value of the applied field using the relationship

$$\theta_0 = \text{acos}\left|\frac{R_f^{AHE}(B_{ext})}{R_{AHE}}\right|. \tag{38}$$

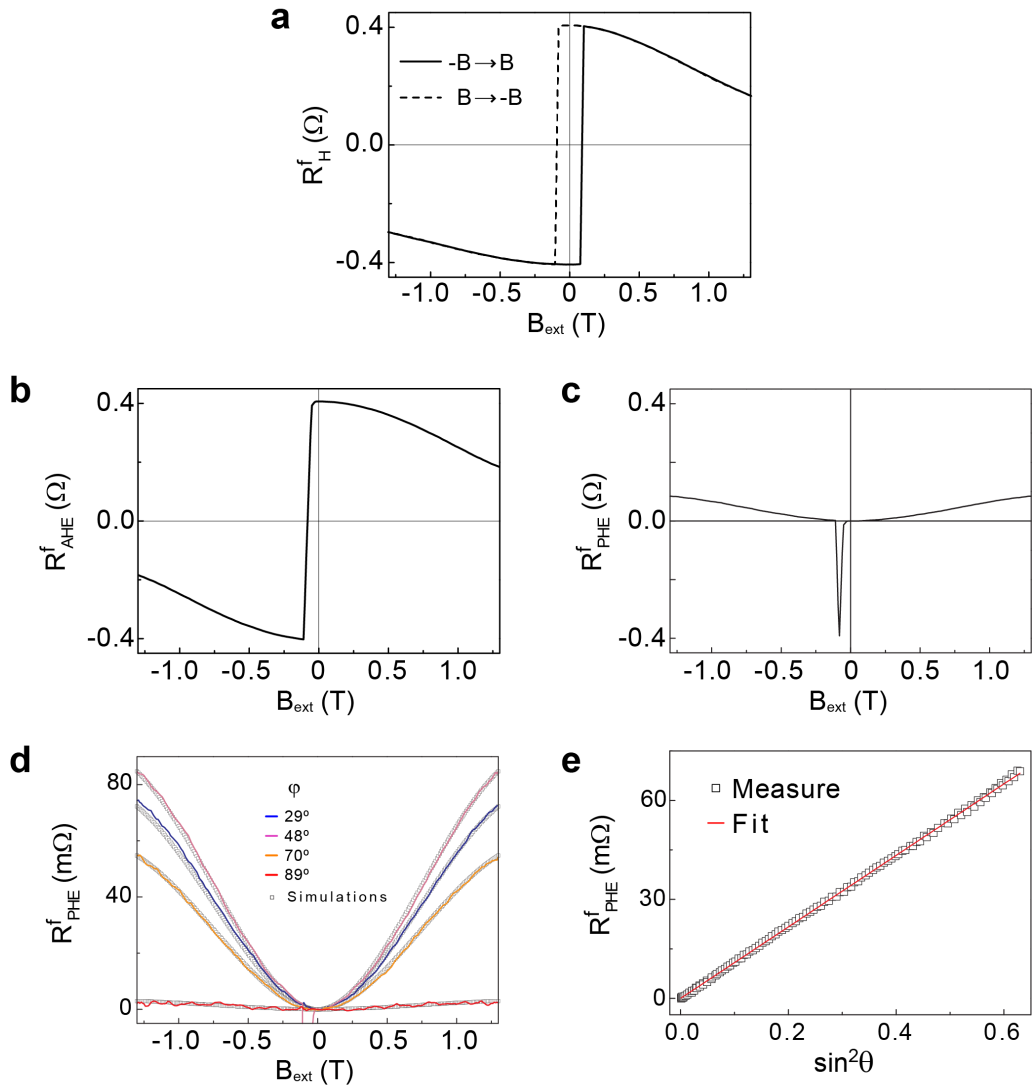

**Figure S3. Separation of AHE and PHE. a,** First harmonic Hall resistance, $R_H^f$, measured at $\theta_B = 80°$, $\varphi = 50°$ and $I = 600$ mA. **b,** Antisymmetric AHE signal, $R_{AHE}^f$. **c,** Symmetric PHE signal, $R_{PHE}^f$. **d,** Comparison between macrospin simulations (dots) and measurements (solid lines) of $R_{PHE}^f$ at different angles $\varphi$. **e,** $R_f^{PHE}$ as a function of $\sin^2 \theta$ showing the expected linear dependence. The slope of this curve gives the PHE resistance, $R_{PHE} = 0.11$ Ω.

## S4. Measurement of an external ac field using the Hall voltage

The measurements of the effective fields based on the harmonic analysis of the Hall voltage described above have been quantitatively checked by applying an ac magnetic field of known amplitude and retrieving its value using Eqs. 37 and 38. To this end, we modulated $B_{ext}$ by a small sinusoidal field of amplitude $B_{ac} = 10.5$ mT at a frequency of 1 Hz. The current frequency was also set to $f = 1$ Hz in order to match the modulation frequency of the electromagnet. The polar component of the ac field acting on the magnetization in this case is given by $b_\theta I + B_{ac}\sin(\theta_B - \theta)$. Since $b_\theta$ and $B_{ac}$ are independent of each other, $R_H^{2f}$ is given by the sum of $R_H^{2f}(b_\theta)$ and $R_H^{2f}(B_{ac})$. By applying the ac field in-phase or out-of-phase with the current, we measure $R_H^{2f}(b_\theta, b_\varphi, B_{ac}) = R_H^{2f}(b_\theta, b_\varphi) + R_H^{2f}(B_{ac})$ and $R_H^{2f}(b_\theta, b_\varphi, -B_{ac}) = R_H^{2f}(b_\theta, b_\varphi) - R_H^{2f}(B_{ac})$, respectively (Fig. S4a). This allows us to separate the two contributions $R_H^{2f}(b_\theta, b_\varphi)$ and $R_H^{2f}(B_{ac})$ by taking the sum and difference of the above relationships. As expected, $R_H^{2f}(b_\theta, b_\varphi)$ coincides with the measurement of $R_H^{2f}$ in the absence of ac field (Fig. S4b). To calculate the amplitude $B_{ac}$ from the experimental data, we use Eq. 37, which leads to

$$B_{ac} = -\frac{2R_H^{2f}(B_{ac})}{(R_{AHE} - 2R_{PHE}\cos\theta_0\sin 2\varphi_0)\frac{d\cos\theta}{dB_{ext}}} \tag{39}$$

The term $\frac{d\cos\theta}{dB_{ext}}$ can be evaluated from the numerical derivative of $\frac{1}{R_{AHE}}\frac{dR_f^{AHE}}{dB_{ext}}$ or, equivalently, from $-2R_H^{2f}(B_{ac})$, as demonstrated experimentally in Fig. S4c. We present three measurements of $R_H^{2f}(B_{ac})$ recorded at $\varphi = 90°$, $60°$, and $0°$ (Fig. S4d). We note that the curves measured at $\varphi = 90°$ and $\varphi = 0°$ have the same shape, whereas the curve measured at $\varphi = 60°$ is asymmetric due to the contribution of the PHE at this angle. Using Eq. 39, we show that we recover correctly the amplitude $B_{ac} = 10.5$ mT independently of $\varphi$ (Fig. S4e), confirming the consistency of the harmonic analysis of the AHE and PHE contributions presented in Sects. 2 and 3. The standard deviation of the data in Fig. S4e is 0.6 mT, which corresponds to a relative error of about 6% on our measurements.

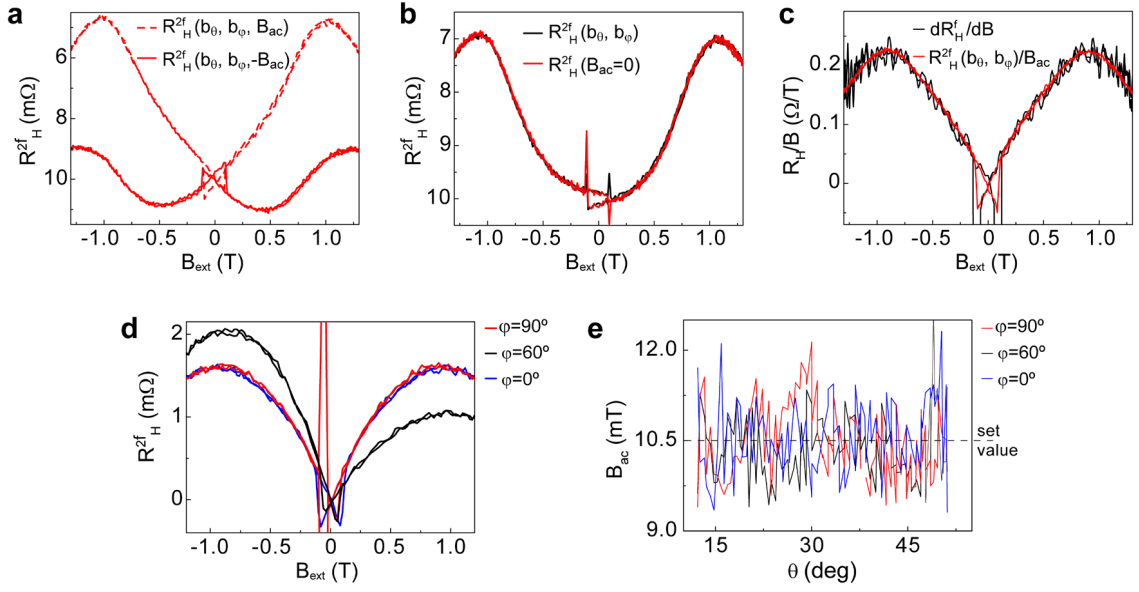

**Figure S4. Measurement of an external ac field using the Hall voltage. a,** $R_H^{2f}$ measured in-phase (dashed line) and antiphase (solid line) with an applied ac magnetic field parallel to **B**$_{ext}$. **b,** Comparison between $R_H^{2f}(b_\theta, b_\varphi) = [R_H^{2f}(b_\theta, b_\varphi, B_{ac}) + R_H^{2f}(b_\theta, b_\varphi, -B_{ac})]/2$ (black line) and $R_H^{2f}(b_\theta, b_\varphi, B_{ac} = 0)$ (red line). **c,** Comparison between the numerical derivative of $R_{AHE}^f$ with respect to the applied field (black line) and $R_H^{2f}(B_{ac}) = [R_H^{2f}(b_\theta, b_\varphi, B_{ac}) - R_H^{2f}(b_\theta, b_\varphi, -B_{ac})]/2$ (red line). **d,** $R_H^{2f}(B_{ac})$ recorded at $\varphi = 90º$, $60º$, and $0º$. **e,** Measured amplitude of the applied ac field as a function of magnetization tilt obtained from the curves shown in **d** using Eqs. 38 and 39.

The external ac field is also useful to determine the direction of **B**$^\perp$ with respect to the Oersted field. In the measurements reported in Fig. S4, **B**$_{ext}$ is perpendicular to the current and, when positive, it is opposite to the Oersted field generated by a positive current flowing in the Pt layer. The Oersted field is determined by the conventional right-hand rule. In Fig. S4a we observe that $R_H^{2f}$ increases (decreases) when $B_{ac}$ is in-phase (out-of-phase) with the current, meaning that $B^{odd}$ and $B_{ac}$ add up for an in-phase measurement, i.e., that **B**$^\perp$ is positive for a positive current. Therefore, we conclude that **B**$^\perp$ is opposite to the Oersted field.

## S5. Measurement of current-induced effective fields in the case of nonzero PHE

Equation 37 relates $R_H^{2f}$ to the effective field components $b_\theta$ and $b_\varphi$. If the PHE is neglected, $b_\theta$ reads

$$b_\theta = \frac{R_H^{2f}\sin(\theta_B - \theta_0)}{IR_{AHE}\left.\dfrac{d\cos\theta}{dB_{ext}}\right|_{\theta_0}}, \qquad (40)$$

which is readily evaluated by noting that

$$R_{AHE}\left.\frac{d\cos\theta}{dB_{ext}}\right|_{\theta_0} = \left.\frac{dR_H^f}{dB_{ext}}\right|_{\theta_0}. \qquad (41)$$

However, despite the fact that the PHE is significantly smaller than the AHE in AlOx/Co/Pt, we find that neglecting the PHE affects the quantitative determination of $b_\theta$ and, therefore, of $B^\perp$ and $B^\parallel$. In fact, Eq. 40 does not allow for a precise calculation of the current-induced fields and one must resort to the more general Eq. 37. In order to solve Eq. 37 for $b_\theta$ and $b_\varphi$, we thus use a recursive procedure that takes advantage of the indipendent measurement of $R_{PHE}$ and $R_{AHE}$ reported in Sect. 3. Starting from $R_H^{2f}$ measured at $\varphi = 0°$ and $90°$ (blue curves in Fig. S5a and b, respectively) we operate as follows:

i) We set $R_{PHE} = 0$ as initial guess and evaluate $b_\theta^0(\varphi = 90°)$ and $b_\varphi^0(\varphi = 0°)$, shown in black in Fig. S5 a and c, respectively.

ii) We set $R_{PHE}$ to its measured value and evaluate $b_\theta^1(\varphi = 90°)$ using the previous estimate of $b_\varphi^0(\varphi = 0°)$. Using both further gives $b_\varphi^1(\varphi = 0°)$.

iii) We evaluate $b_\theta^1(\varphi = 0°)$ using $b_\varphi^1(\varphi = 0°)$, which further gives $b_\varphi^1(\varphi = 90°)$, as shown in Fig. S5 b,d, blue curves.

iv) Steps ii) and iii) are repeated until we achieve convergence (red curves in Fig. S5).

Figures S5c and d show the successive iterations that lead to the final form of $b_\theta$ at $\varphi = 90°$ and $0°$, respectively. This procedure is independent on the choice of coefficients used to represent $b_\theta$ and $b_\varphi$.

We note that Eqs. 24-27 imply that $b_\theta = B^\perp$ at $\varphi = 90°$ and $b_\theta = B^\parallel$ at $\varphi = 0°$. Recalling that by truncating the angular expansion of the torques to fourth order we have

$$Ib_\theta(\varphi = 0°) = T_0^\parallel + T_2^\parallel \sin^2\theta + T_4^\parallel \sin^4\theta, \tag{42}$$

$$Ib_\theta(\varphi = 90°) = -\cos\theta \, (T_0^\perp + T_2^\perp \sin^2\theta + T_4^\perp \sin^4\theta), \tag{43}$$

$$Ib_\varphi(\varphi = 0°) = -T_0^\perp, \tag{44}$$

$$Ib_\varphi(\varphi = 90°) = -\cos\theta \, T_0^\parallel, \tag{45}$$

the values of the coefficients $T_{0,2,4}^\parallel$ and $T_{0,2,4}^\perp$ are obtained by fitting $b_\theta(\varphi = 0°)$ and $b_\theta(\varphi = 90°)$ using Eqs. 42 and 43.

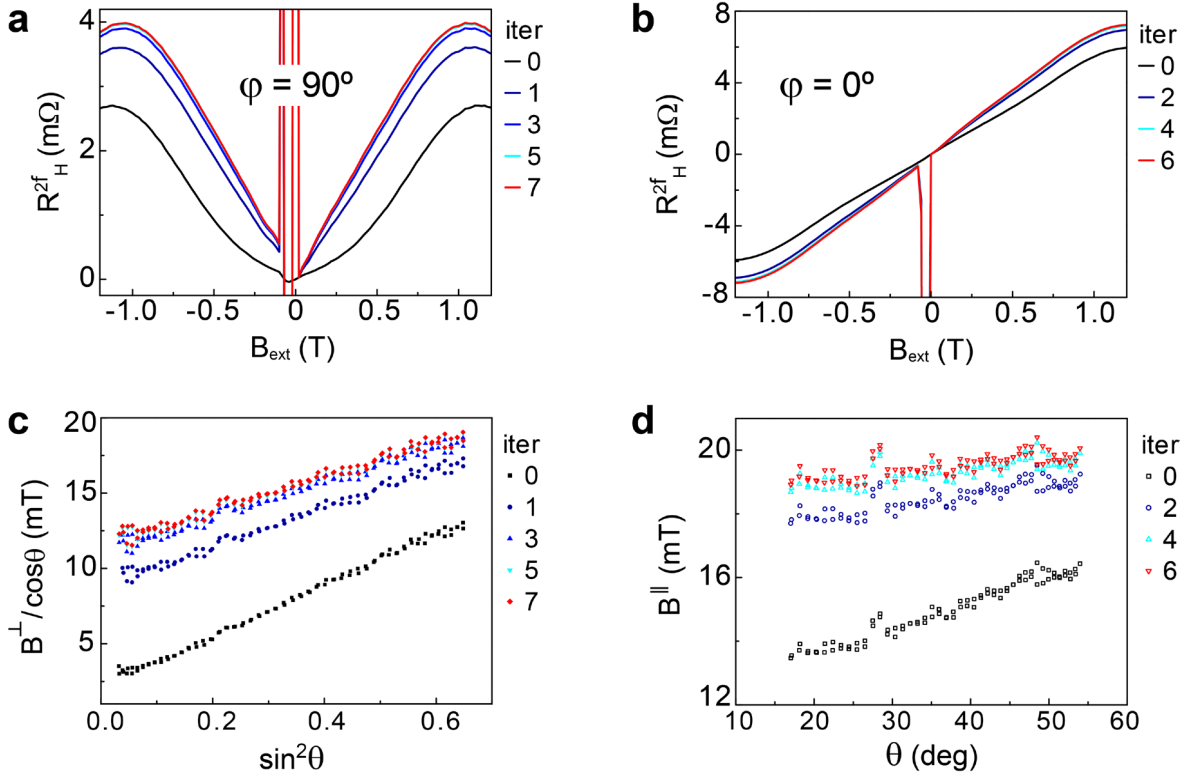

**Figure S5. Recursive procedure to determine $B^\perp$ and $B^\parallel$ in the case of nonzero PHE. a,** $R_H^{2f}$ measured at $\varphi = 90°$ (black line) and successive iterations that lead to the separation of the pure AHE contribution $R_{AHE}^{2f}(\varphi = 90°)$ (red line). **b,** $R_H^{2f}$ measured at $\varphi = 0°$ (black line) and successive iterations that lead to $R_{AHE}^{2f}(\varphi = 0°)$ (red line). **c,** Iterations leading to convergent $B^\perp$ values at $\varphi = 0°$. **d,** Iterations leading to convergent $B^\parallel$ values at $\varphi = 0°$.

## S6. Macrospin simulations

To validate the iteration procedure described above, we performed numerical simulations using a macrospin model that includes the effects of SOTs. The model allows us to simulate $R_H^f$ and $R_H^{2f}$ as a function of $B_{ext}$ starting from the equilibrium equation for the magnetization:

$$\boldsymbol{m} \times \boldsymbol{B}_{ext} + \boldsymbol{m} \times \boldsymbol{B}_k + \boldsymbol{T}^\perp + \boldsymbol{T}^\| = 0, \qquad (46)$$

where the first term represents the torque due to the external field and the second term the torque due to the anisotropy field $\boldsymbol{B}_k = [B_{k1}(\boldsymbol{m} \cdot \boldsymbol{z}) + B_{k2}(\boldsymbol{m} \cdot \boldsymbol{z})^3]\boldsymbol{z}$, with second and fourth order anisotropy constants $B_{k1}$ = -0.95 T and $B_{k2}$ = -0.2 T, respectively. We assume here the simplest form of $\boldsymbol{T}^\perp$ and $\boldsymbol{T}^\|$ compatible with the experimental results of $AlO_x$/Co/Pt, namely

$$\boldsymbol{T}^\perp = \cos\phi\, T_0^\perp \mathbf{e}_\theta - \cos\theta \sin\phi\, (T_0^\perp + T_2^\perp \sin^2\theta)\mathbf{e}_\phi \qquad (47)$$

and

$$\boldsymbol{T}^\| = \cos\theta \sin\phi\, T_0^\| \mathbf{e}_\theta + \cos\phi\, T_0^\| \mathbf{e}_\phi \qquad (48)$$

and set the SOT coefficients to $T_0^\perp$ = -12.0 mT, $T_2^\perp$ = -11.0 mT, and $T_0^\|$ = 19.0 mT. These coefficients are modulated by the current, which is proportional to $e^{i2\pi ft}$. By solving Eq. 43 at each instant $t$ for the angles $\theta_0$ and $\varphi_0$ that define the equilibrium direction of $\boldsymbol{m}$ and taking $R_{AHE}$ = 0.8 Ω and $R_{PHE}$ = 0.09 Ω, we calculate $V_H(t)$ using Eq. 29. Then, $V_H(t)$ is Fourier-transformed to obtain $R_H^f$ and $R_H^{2f}$ as a function of $B_{ext}$. Figure S6a and b show that the simulations faithfully reproduce the shape of the experimental $R_H^{2f}$ curves measured at $\varphi$ = 0° and 90°, respectively (see Fig. 2, main text). Further, by applying the iteration steps i-iv described above to the simulated data (Fig. S6c-f), we find a very similar convergence behaviour to that reported for the experimental data in Fig. S5. Moreover, after a maximum of seven iteration steps, we recover the values $T_0^\perp$ = -11.99 mT, $T_2^\perp$ = -11.01 mT, and $T_0^\|$ = 18.97 mT, thus confirming the validity of our analysis.

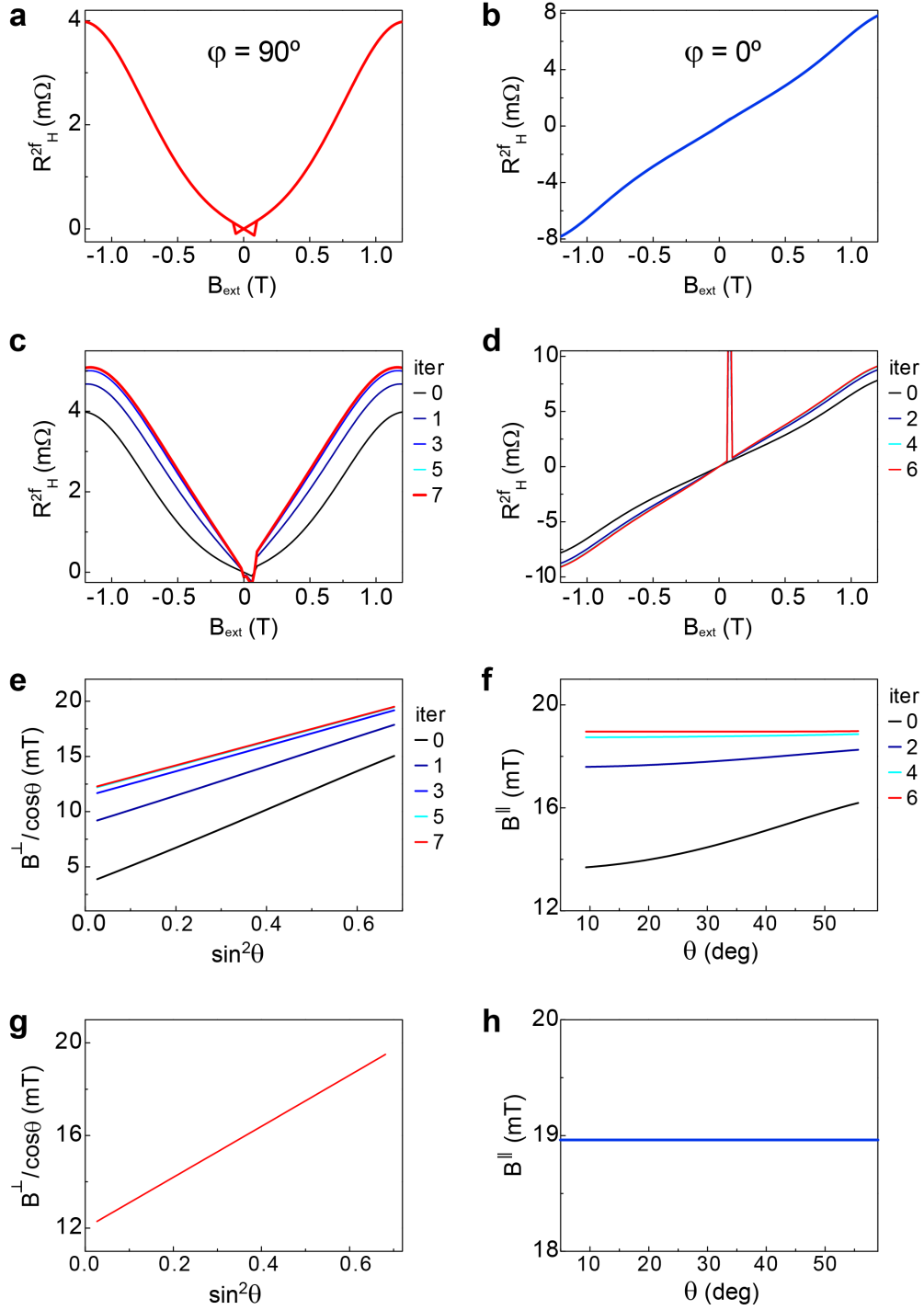

**Figure S6. Macrospin simulations of $R_H^{2f}$, $B^\perp$, and $B^\parallel$ including both the AHE and PHE contributions to the Hall voltage. a,** Simulated $R_H^{2f}$ signal at $\varphi = 90°$ and **b,** $\varphi = 0°$ (see text for details). **c, d,** Iteration procedure to separate the pure AHE components (**c**) $R_{AHE}^{2f}(\varphi = 90°)$ and (**d**) $R_{AHE}^{2f}(\varphi = 0°)$ applied to the simulated signal. Iterative estimate of **e,** $B^\perp$ at $\varphi = 0°$ and **f,** $B^\parallel$ at $\varphi = 0°$. Final result for **g,** $B^\perp$ and **h,** $B^\parallel$ derived from the macrospin simulation of the Hall voltage and analysis of the second harmonic AHE and PHE. $B^\perp$ and $B^\parallel$ coincide with the input values assumed in the simulations.

## S7. Sample-dependent offset and Nernst-Ettingshausen effect

Aside from the AHE and PHE, the measurement of $R_H^{2f}$ is sensitive to sample-dependent contributions to the Hall voltage, namely the misalignment of the voltage leads and thermoelectric effects.[2] The current flowing into the Hall cross can generate a temperature gradient due to inhomogeneous heating in correspondence of fabrication defects, generally on the corners of the sample. This gradient induces two types of thermoelectric voltages, the Seebeck effect and the Anomalous Nernst-Ettingshausen effect (ANE). Since heating is modulated by the ac current, we detect both contributions in the *2f* component of the Hall voltage. We note that these effects vary in amplitude and sign from sample to sample and that, once accounted for, $R_H^{2f}$ presents the same magnetization dependence in all samples.

The largest sample-dependent contribution to $R_H^{2f}$ is a constant offset ($R_{Offset}$) due to the asymmetry of the voltage probes as well as to the Seebeck effect. The ANE, on the other hand, gives a contribution to $R_H^{2f}$ that is magnetization dependent and induces a small asymmetry in the raw $R_H^{2f}$ curves. The voltage induced by ANE is perpendicular to the temperature gradient and $m_z$, mimicking the AHE with much smaller amplitude. (Fig. S7a and b). Both the offset and ANE contributions to $R_H^{2f}$ can be determined by performing a measurement with $\mathbf{B}_{ext}$ // $\mathbf{z}$, as shown in Fig. S7a. In this case, the SOT and Oersted fields do not contribute to the second harmonic signal since the variation of the Hall resistance is symmetric with respect to θ=0°. Hence, the residual $R_H^{2f}$ signal is related uniquely to $R_{Offset}$ and the ANE. As expected for the ANE, $R_H^{2f}$ is hysteretic and has the same field dependence of $R_H^f$, which is proportional to $m_z$. One can easily deduce the amplitude of the ANE, $R_{ANE}$, by taking the difference between the extrema of $R_H^{2f}$ measured for $\mathbf{B}_{ext}$ // $\mathbf{z}$. We find $R_{ANE} = 0.1$ mΩ for a current of 635 μA, which is comparable with other values found in the literature.[3] For arbitrary orientation of $\mathbf{B}_{ext}$, we find that $R_{ANE}$ coincides with the difference of $R_H^{2f}$ measured at zero field for positive and negative sweeps of $B_{ext}$, whereas their average gives $R_{Offset}$. Finally, both $R_{Offset}$ and $R_{ANE}$ can be subtracted from the raw data, giving:

$$R_H^{2f} = R_{raw}^{2f} - R_{ANE}\frac{R_H^f}{2R_{AHE}} - R_{Offset} \ . \tag{49}$$

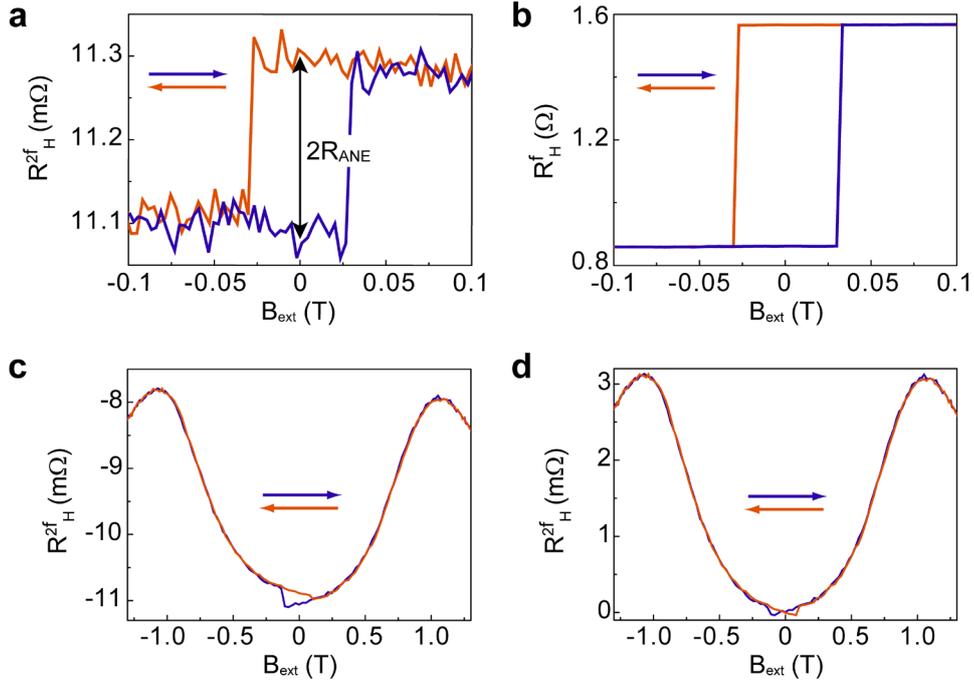

**Figure S7. Measurement and subtraction of the Anomalous Nernst effect. a,** $R_H^{2f}$ measured at $\theta_B = 0°$, $\varphi = 90°$, and $I = 635$ μA. **b,** $R_H^f$ simultaneously measured with **a**. **c, d,** $R_H^{2f}$ measured at $\theta_B = 85°$, $\varphi = 90°$, and $I = 635$ μA before (**c**) and after (**d**) subtraction of the ANE and constant offset.

Figures S7c and d show $R_H^{2f}$ before and after subtraction of $R_{Offset}$ and $\Delta R_{ANE}$, respectively.

**S8. Current dispersion in the Hall voltage probes**

The SOT/current ratios reported in the main text are calculated without taking into account the spread of the charge current into the voltage leads of the Hall cross. Depending on the relative width of the Hall voltage probes with respect to the current probes, the actual current giving rise to the SOTs may be smaller than the total current injected into the device. Numerical simulations of the current flow in a Hall cross show that the current density in the central region of the cross reduces by up to 23% (8%) in junctions where the width of the voltage probes is equal (half) the width of the current probes.[4] The dimensions of the Hall cross in our devices are 500 x 500 nm$^2$, 1000 x 1000 nm$^2$, and 1000 x 500 nm$^2$. Figure 3 in the main text shows that the SOT/current ratios measured in the 500 x 500 nm$^2$ and 1000 x 1000 nm$^2$ devices are similar, whereas in the 1000 x 500 nm$^2$ device they are larger by 20-30% relative to the symmetric probes. Figure S8 shows a set of measurements for a 1000 x 500 nm$^2$ device of AlO$_x$/Co/Pt recorded at $\varphi = 90°$, $0°$, and $\theta_B = 82°$. For an ac current amplitude set to $I = 1040$ μA, corresponding to a current density $j = 2.91 \times 10^7$ A/cm$^2$, we obtain $T_0^\perp = -16 \pm 1$ mT, $T_2^\perp = -12.8 \pm 0.8$ mT, and $T_0^\parallel = 25 \pm 1$ mT. For a 1000 x 1000 nm$^2$ device patterned on the same wafer, we measured $T_0^\perp = -11.4$ mT, $T_2^\perp = -9.3$ mT, and $T_0^\parallel = 18.7$ mT for the same current amplitude. We attribute this effect to the larger current density flowing in the magnetic region of the 1000 x 500 nm² device due to the reduced width of the voltage probes.

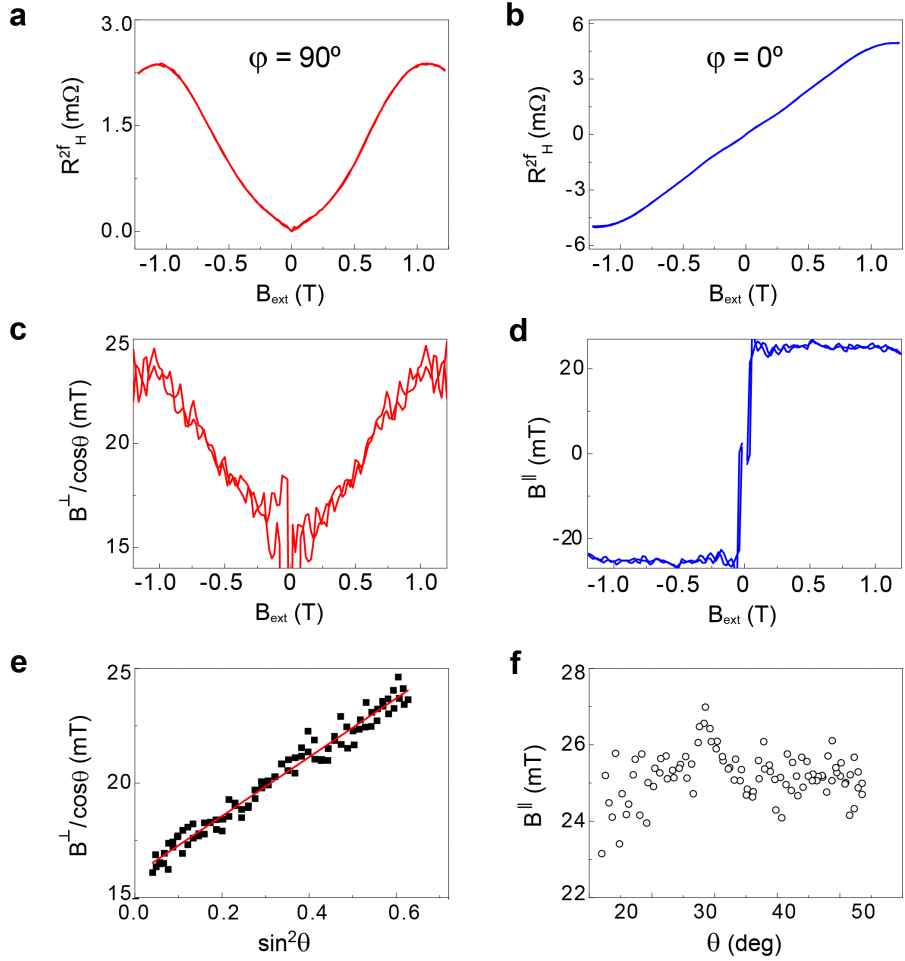

**Figure S8. Second harmonic Hall resistance and current-induced spin-orbit fields measured on a 1000 x 500 nm² device. a,** $R_H^{2f}$ as a function of $B_{ext}$ applied at $\theta_B = 82°$, $\varphi = 90°$ and **b,** $\theta_B = 82°$, $\varphi = 0°$. The amplitude of the ac current is 1.136 mA. **c,** Effective field $B^\perp/\cos\theta$ as a function of $B_{ext}$ ($\varphi = 90°$). **d,** Effective field $B^{/\!/}$ as a function of $B_{ext}$ ($\varphi = 0°$). **e,** $B^\perp/\cos\theta$ as a function of $\sin^2\theta$ measured at $\varphi = 90°$. The solid line is a fit to the function $T_0^\perp + T_2^\perp \sin^2\theta$ with $T_0^\perp = -16.0$ mT and $T_2^\perp = -12.8$ mT. **f,** $B^{/\!/}$ as a function of $\theta$ measured at $\varphi = 90°$.

## S9. Measurements in the case of nonuniform magnetization

Our measurements are based on detecting small current-induced oscillations of the magnetization about its equilibrium direction, which is determined by the anisotropy field and $\mathbf{B}_{ext}$. The magnetization curves ($R_H^f$) measured for $\theta_B \leq 85°$ show reversible behavior beyond the switching field, consistently with coherent rotation of the magnetization towards $\mathbf{B}_{ext}$. At $\theta_B > 85°$, however, we observe irreversible jumps of the Hall resistance due to the formation of magnetic domains. These jumps are also detected in the $R_H^{2f}$ curves, as shown in Fig. S9 for a geometry $\theta_B = 87°$, $\varphi = 90°$ and $I = 635$ µA. For this reason, the measurements reported in this paper are limited to $\theta_B \leq 85°$.

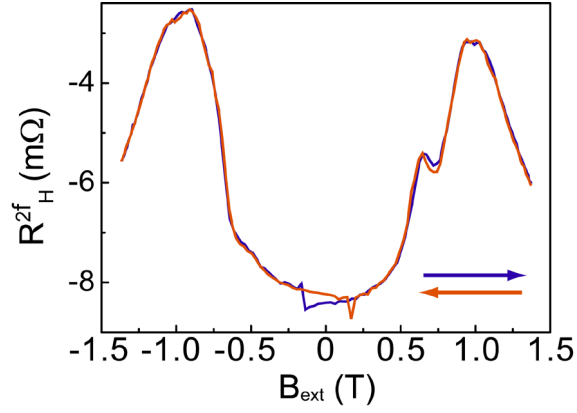

**Figure S9. Second harmonic Hall resistance in the presence of magnetic domain nucleation. a,** $R_H^{2f}$ as a function of $B_{ext}$ applied at $\theta_B = 87°$, $\varphi = 90°$. The amplitude of the ac current is 635 mA. A jump of the signal is observed between 0.5 and 1 T.

## S10. Comparison of AC and DC detection methods

We present here a comparison of our AC detection method with DC measurements of the Hall voltage, analogue to those performed by Liu et al. in Refs. 5 and 6. These authors considered a scalar model where the torques due to the external field, magnetic anisotropy, and current are collinear. In the macrospin approximation, this leads to the following torque equation at equilibrium:

$$B_{ext} \sin(\theta_0 - \theta_H) - B_k \sin\theta_0 \cos\theta_0 + T(I) = 0. \qquad (50)$$

By estimating the equilibrium magnetization angle $\theta_0$ using the AHE (Eq. 38), one can define two magnetic field values, $B_+(\theta_0)$ and $B_-(\theta_0)$, as the value of $B_{ext}$ required to produce a given value of $\theta_0$ for positive and negative current, respectively.[5] From Eq. 50 one has $B_\pm(\theta_0) = [B_k \sin\theta_0 \cos\theta_0 \mp T(|I|)]/\sin(\theta_0 - \theta_H)$ and, finally,

$$T(I) = \frac{B_-(\theta_0) - B_+(\theta_0)}{2} \sin(\theta_0 - \theta_H). \qquad (51)$$

In practice, $B_\pm(\theta_0)$ are calculated by measuring the Hall resistances for positive and negative current $[R_H(I_\pm)]$ as a function of applied field.

By assuming the simplest form for the torques, $\boldsymbol{T}^{\parallel} = T_0^{\parallel} \boldsymbol{m} \times (\boldsymbol{y} \times \boldsymbol{m})$ and $\boldsymbol{T}^{\perp} = T_0^{\perp}(\boldsymbol{y} \times \boldsymbol{m})$, Liu et al. used Eq. 51 to measure $T_0^{\parallel}$ and $T_0^{\perp}$ for $B_{ext}$ applied in the *xz* and *yz* plane, respectively (note that, with respect to our notation, the *x* and *y* axis are interchanged in Refs. 5 and 6). For AlO$_x$/Co(0.6 nm)/Pt(2 nm) annealed in vacuum at 350 C, Liu et al. concluded that $T_0^{\parallel}$ = 0.33±0.06 mT/mA (1.7 mT per $10^7$ A/cm$^2$) and that $T_0^{\perp}$ = 0, within the sensitivity of the experiment (1.3 mT per $10^7$ A/cm$^2$). Because the SOT amplitudes are generally very sensitive to the sample growth details, it is not surprising that we obtain different torque values, at least for $T_0^{\parallel}$. However, Eq. 51 assumes that the magnetization remains confined in the plane defined by the external field and *z* axis, which is not true if both $\boldsymbol{T}^{\parallel}$ and $\boldsymbol{T}^{\perp}$ are present, as can be seen from Eqs. 18, 19 and 24, 25 (see also Fig. 11b). Moreover, if the magnetization deviates from the *xz* or the *yz* plane, the PHE contributes to $R_H$ in a way that is not symmetric for positive and negative current. These effects, as well as the presence of additional torque components besides $T_0^\parallel$ and $T_0^\perp$, have not been taken into account in the analysis of Refs. 5 and 6.

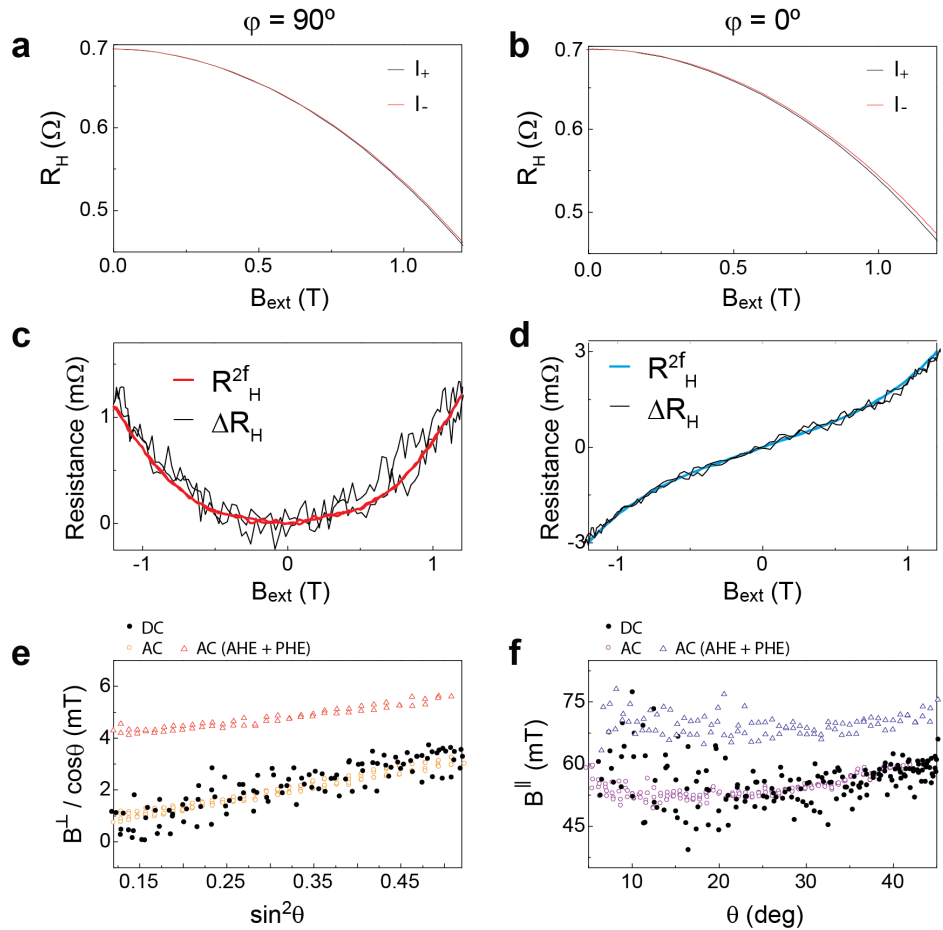

**Figure S10. Comparison between AC and DC detection methods.** $R_H$ measured on a Hall cross of 3000 nm (current injection) by 500 nm (Hall voltage) for $I_\pm = \pm 1.2$ mA, $\theta_B = 86°$, and **a,** $\varphi = 90°$, **b,** $\varphi = 0°$. **c, d,** Comparison of $R_H^{2f}$ and $\Delta R_H/2$. **e,** Comparison of $B^{\perp}$ extracted from $R_H^{2f}$ (open symbols) and $\Delta R_H$ (filled symbols). **f,** Same for $B^{\parallel}$. The data corrected for the PHE are also shown. The integration time used in the AC method was 10s. In the DC method, each field point has been averaged for 10 s.

Here, we show that, if the PHE is neglected, the two methods give equivalent results, but that the AC measurements give a better signal-to-noise ratio compared to DC, as expected. Figures S10a and b show $R_H(I_+)$ and $R_H(I_-)$ measured for $B_{ext}$ applied in the *yz* ($\varphi = 90°$) and *xz* ($\varphi = 0°$) plane, respectively. In Fig. S10c and d we compare $R_H^{2f}$ with the corresponding quantity in a DC measurement, $\Delta R_H/2 = [R_H(I_+) - R_H(I_-)]/2$. The fields $B_+(\theta_0)$ and $B_-(\theta_0)$ introduced above can be easily derived from $\Delta R_H$. We observe that, apart from the noise level, $R_H^{2f}$ and $\Delta R_H$ present the same dependence on the external field. Figures S10e and f show that, if $R_{PHE}$ is set to zero, the current-induced fields $B^\perp$ and $B^\parallel$ obtained from the analysis of $R_H^{2f}$ and $\Delta R_H$ using Eqs. 40 and 41 give similar results, although the scattering of the DC data is much larger. The AC data corrected for the PHE are also shown. Clearly, the AC method allows for more sensitive measurements and, therefore, to work in a low current regime where the magnetization behaves coherently and thermal effects are small.

## S11. Dynamic simulations of the $m_y$ component generated by $T^\parallel$

As mentioned above and in the main text, a remarkable difference between our data and those reported by Liu et al. is the null result obtained for $T^\perp$ in Ref. 5. We pointed out the reduced sensitivity of DC measurements and differences in sample preparation as possible clues for such a discrepancy. On the other hand, Liu et al. confuted the interpretation of previous measurements of $T^\perp$ by noting that the spin transfer torque $T^\parallel$ due to the spin Hall effect may induced a tilt of the magnetization along the *y* axis, similar to that expected by $T^\perp$ (Ref. 5). This would be the case if $|T_0^\parallel| > B_k/2$, that is, if $T_0^\parallel$ is so large as to overcome the anisotropy field and induce switching. However, the method presented in this work is based on small oscillations of the magnetization and the torque amplitudes have been measured in the low current regime (Fig. 4 main text), which is very far from the regime $|T_0^\parallel| > B_k/2$ hypothized in Ref. 5. We have carried out dynamic simulations of the *x*, *y*, *z* magnetization components subject to either $T^\perp + T^\parallel$ or $T^\parallel$ alone to confirm this point, by numerically solving the LLG equation $\dot{\boldsymbol{m}} = -|\gamma_0|[\boldsymbol{m} \times (\boldsymbol{B}_{ext} + \boldsymbol{B}_k) + \boldsymbol{T}^\parallel + \boldsymbol{T}^\perp] + \alpha\, \boldsymbol{m} \times \dot{\boldsymbol{m}}$, where $\gamma_0$ is the gyromagnetic ratio and $\alpha$ the Gilbert damping. The simulations were performed using the following parameters: $B_k = 1$ T, $\alpha = 0.5$, and $B_{ext}= 0.1$ T, $\varphi=0°$, $\theta_H=90°$. Figure S11a shows that for $T_0^\perp = T_2^\perp = 0$ and $T_0^\parallel = 0.3$ T $< B_k/2$, no magnetization component appears along the $y$ axis. A nonzero $m_y$ only appears if $T^\perp$ is turned on, as in Fig. S11b or when $T_0^\parallel = 0.6$ T $> B_k/2$, as in Fig. S11c.

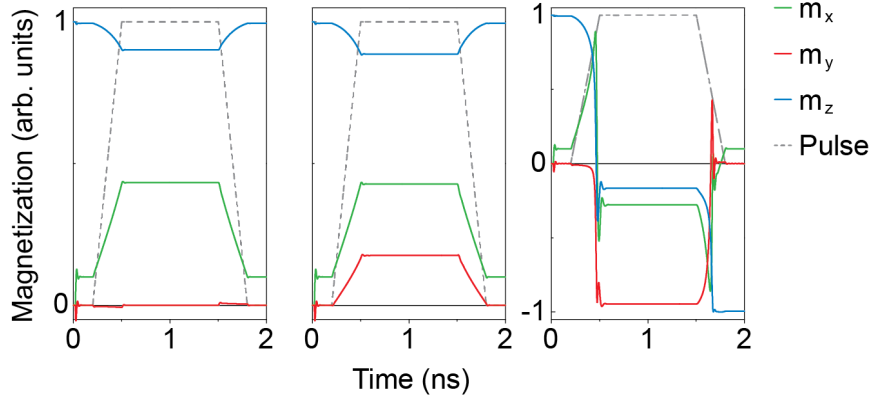

**Figure S11. Dynamic simulations of $m_x$, $m_y$, and $m_z$ for different torque amplitudes. a,** $T_0^\perp = T_2^\perp = 0$ and $T_0^\parallel = 0.3$ T, **b,** or $T_0^\perp = T_2^\perp = 0.15$ T and $T_0^\parallel = 0.3$ T **c,** $T_0^\perp = T_2^\perp = 0$ and $T_0^\parallel = 0.6$ T. The dashed line represents the injected current.

## S12. Influence of thermal effects

It is well known that heating reduces the magnetic anisotropy field and saturation magnetization of thin films. Therefore ohmic heating induced by the current effectively softens the magnetization of the ferromagnetic layer, increasing the susceptibility of the magnetization to the spin torques. A consequence of this effect is to introduce a spurious increase of the torque amplitudes measured at high current, which may add itself to the intrinsic temperature dependence of the torques. However, heating effects are proportional to $I^2$ and hence should appear as an odd harmonic component of the Hall resistance, $R_H^{3f}$, whereas $R_H^{2f}$ should reflect the presence of spin torque components that are linear with the current. To check the validity of this statement, we implemented Joule heating in our macrospin simulations and compared them with experimental data obtained for two different current amplitudes (Fig. S12). Heating was modeled by considering a reduced anisotropy field $B_k(1 - a_k I^2)$ and saturation magnetization $M_s(1 - a_M I^2)$, using the following parameters: $B_k = 1$ T, $\mu_0 M_s = 1$ T, $R_{AHE}= 0.8$ Ω, $R_{PHE}=$ 0.1 Ω, $T_0^\perp = -0.12$ T, $T_2^\perp = -0.11$ T, and $T_0^\parallel = 0.19$ T. The coefficients $a_k=0.14$ and $a_M=0.035$ were derived from the measured current dependence of $B_k$ and $R_{AHE}$, respectively. We performed simulations of $R_H^f$, $R_H^{2f}$, and $R_H^{3f}$, as shown in Fig. S12. In the top panels, we observe a small decrease of the $R_H^f$ amplitude at remanence, due to the decrease of $M_s$, and at high field due to the reduction of $B_k$. Both the experiments and simulations show a clear $R_H^{3f}$ signal (bottom panels) that disappears when heating is turned off in the simulations (blue line). On the other hand, the measurements of $R_H^{2f}$ normalized by the current amplitude nearly superpose (middle panels), meaning that heating effects have only a minor influence on the spin torque measurements in this current regime. This is in agreement with the current dependence of the torque coefficients reported in Fig. 4 of the main text, which is linear below $j = 1.5 \times 10^7$ A/cm$^2$. Above this limit, our simulations show that $R_H^{2f}$ gradually increases due to the reduction of $M_s$, whereas $R_H^{3f}$ is affected by $M_s$ (hysteretic part) and $B_k$ (high-field part). Finally, we simulated the case where $T^\perp$ is set to zero to check whether a heat-induced modulation of the anisotropy field can mimic the action of the field-like torque, as suggested in Ref. 5. This turns out not to be the case, since $R_H^{2f}$ is zero in such a case (green line).

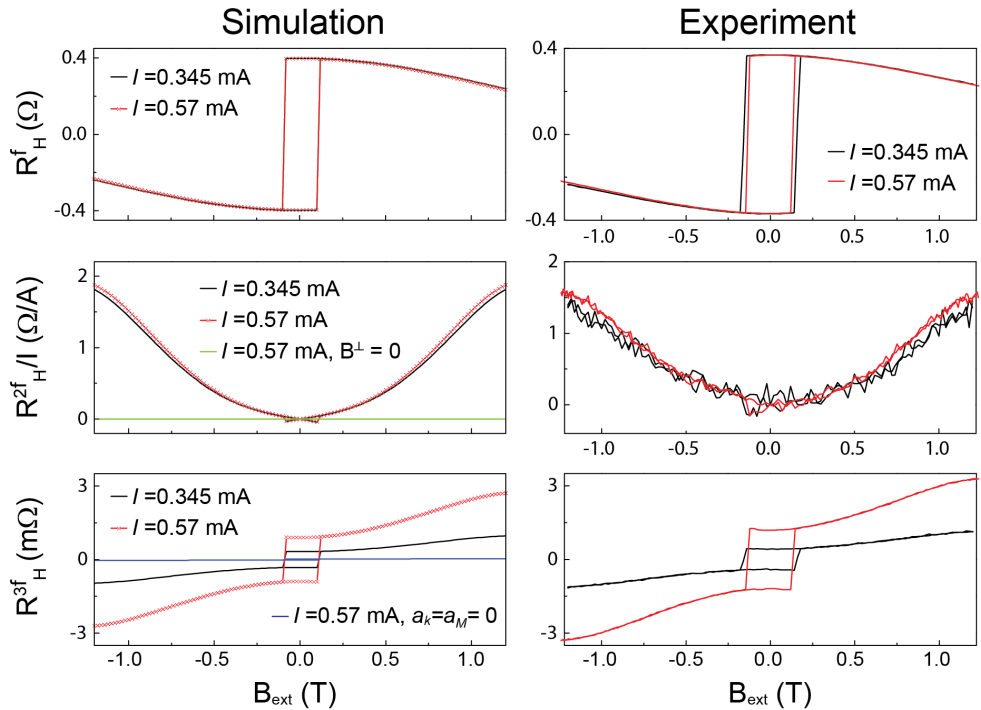

**Figure S12. Macrospin simulations of thermal induced effects and comparison with experimental data.** Simulations (left column) and measurements (right column) of $R_H^f$, $R_H^{2f}$, and $R_H^{3f}$ for $\varphi = 90°$. The measurements were performed on a 1000 x 1000 nm$^2$ AlO$_x$/Co/Pt device. See text for details.

## S13. Supplementary references

[1] Pauyac, C. O., Wang, X., Chshiev, M. & Manchon, A. Angular dependence and symmetry of Rashba spin torque in ferromagnetic heterostructures, *http://arxiv.org/pdf/1304.4823*.

[2] Lindberg, O. Hall Effect. *Proc. Inst. Radio Engrs*. **40**, 1414-1419 (1952).

[3] Seki, T. et al., Giant spin Hall effect in perpendicularly spin-polarized FePt/Au devices, *Nat. Mater.* **7**, 125-129 (2008).

[4] Ibrahim, I. S., Schweigert, V. A. & Peeters, F. M. Diffusive transport in a Hall junction with a microinhomogeneous magnetic field. *Phys. Rev. B* **57**, 15416-15427 (1998).

[5] Liu, L., Lee, O. J., Gudmundsen, T. J., Ralph, D. C. & Buhrman, R. A. Current-Induced Switching of Perpendicularly Magnetized Magnetic Layers Using Spin Torque from the Spin Hall Effect. *Phys. Rev. Lett.* **109**, 096602 (2012).

[6] Liu, L., Pai, C. F., Li, Y., Tseng, H. W., Ralph, D. C. & Buhrman, R. A. Spin-torque switching with the giant spin Hall effect of tantalum. *Science* **336**, 555-558 (2012).